\documentclass[11pt]{article}

\PassOptionsToPackage{table,dvipsnames}{xcolor}
\usepackage[final]{acl}

\usepackage{times}
\usepackage{amsmath}
\usepackage{latexsym}
\usepackage{booktabs}
\usepackage{multirow}
\usepackage{makecell}
\usepackage{amssymb}
\usepackage{subcaption}
\usepackage{algorithm}
\usepackage{algpseudocode}
\usepackage{enumitem}

\usepackage[T1]{fontenc}

\usepackage[utf8]{inputenc}

\usepackage{microtype}

\usepackage{inconsolata}

\usepackage{graphicx}

\usepackage{tcolorbox}
\tcbuselibrary{breakable}

\usepackage{tikz}
\usetikzlibrary{arrows.meta}
\usepackage{listings}
\usepackage{tcolorbox}
\tcbuselibrary{breakable}
\PassOptionsToPackage{table,dvipsnames}{xcolor}
\usetikzlibrary{arrows.meta,calc,backgrounds,fit}
\usepackage{caption}
\usepackage{multicol}

\usepackage{xcolor}
\usepackage{listings}
\usepackage{tcolorbox}  \tcbuselibrary{breakable,skins}
\usepackage{caption}
\definecolor{pathwayHL} {RGB}{186,225,255}
\definecolor{corrHL}    {RGB}{255,193,193}
\definecolor{threshHL}  {RGB}{180,240,180}
\definecolor{footnoteHL}{RGB}{255,243,166}
\definecolor{anchorHL}  {RGB}{225,195,255}
\definecolor{enrichHL}  {RGB}{255,218,185}
\newcommand{\lsfont}{\ttfamily\fontsize{6.5pt}{8pt}\selectfont}
\newcommand{\hlP}[1]{\colorbox{pathwayHL}{\lsfont#1}}
\newcommand{\hlC}[1]{\colorbox{corrHL}{\lsfont#1}}
\newcommand{\hlT}[1]{\colorbox{threshHL}{\lsfont#1}}
\newcommand{\hlF}[1]{\colorbox{footnoteHL}{\lsfont#1}}
\newcommand{\hlA}[1]{\colorbox{anchorHL}{\lsfont#1}}
\newcommand{\hlE}[1]{\colorbox{enrichHL}{\lsfont#1}}

\usepackage{adjustbox}
\usepackage{tabularx}
\usepackage{siunitx}
\usepackage{pifont}
\usepackage{tcolorbox}
\usepackage{listings}
\lstdefinelanguage{json}{
    basicstyle=\ttfamily\small,
    showstringspaces=false,
    breaklines=true,
    frame=none,
    morecomment=[l]{//},
    commentstyle=\color{OliveGreen}\itshape,
    stringstyle=\color{BrickRed},
    morestring=[b]",
    literate=
     *{0}{{{\color{purple}0}}}{1}
      {1}{{{\color{purple}1}}}{1}
      {2}{{{\color{purple}2}}}{1}
      {3}{{{\color{purple}3}}}{1}
      {4}{{{\color{purple}4}}}{1}
      {5}{{{\color{purple}5}}}{1}
      {6}{{{\color{purple}6}}}{1}
      {7}{{{\color{purple}7}}}{1}
      {8}{{{\color{purple}8}}}{1}
      {9}{{{\color{purple}9}}}{1},
}

\newcommand{\dataset}{\textsc{EU-TaxoStruct}}
\newcommand{\framework}{\textsc{RegReAct}}
\newcommand{\odr}{\textsc{ODR}}
\newcommand{\cmark}{\ding{51}}
\newcommand{\xmark}{\ding{55}}

\newcommand{\apxline}[7][]{%
  \par\noindent
  \begin{minipage}[b]{\dimexpr\textwidth-#3\relax}
    \hspace{#3}%
    \makebox[#4][l]{#5}%
    \ifx&#1&%
      \textbf{#6}\dotfill\makebox[2em][r]{#7}%
    \else
      \hyperref[#1]{\textbf{#6}\dotfill\makebox[2em][r]{#7}}%
    \fi
  \end{minipage}\vspace{0.2em}
}

\title{REGREACT: Self-Correcting Multi-Agent Pipelines for Structured Regulatory Information Extraction}

\author{
  Mohammed Ali \quad Abdelrahman Abdallah \quad Adam Jatowt \\
  University of Innsbruck, Austria \\
  \texttt{\{mohammed.ali, abdelrahman.abdallah, adam.jatowt\}@uibk.ac.at}
}

\begin{document}
\maketitle

\begin{abstract}
Extracting structured, machine-readable compliance criteria from regulatory documents remains an open challenge.
Single-pass language models hallucinate structural elements, lose hierarchical relationships, and fail to resolve inter-document dependencies.
We introduce \textsc{RegReAct}, a self-correcting multi-agent framework that decomposes regulatory information extraction into seven specialized stages, each with an \textit{Observe--Diagnose--Repair} (ODR) loop that validates outputs against the source, correcting not only model hallucinations but also cross-reference errors in the regulations themselves.
To ensure structural accuracy, \textsc{RegReAct} constructs a typed criterion graph; to ensure completeness, it resolves external dependencies by retrieving, summarizing, and embedding referenced legal content inline, producing self-contained outputs.
Applying \textsc{RegReAct} to three EU Taxonomy Delegated Acts, we construct a dataset comprising 242 activities with over 4,800 hierarchical criteria, thresholds, and enriched source summaries.
Evaluation against a GPT-4o single-pass baseline shows that \textsc{RegReAct} outperforms it across all structural and semantic metrics \footnote{\url{https://github.com/DataScienceUIBK/REGREACT}}.
\end{abstract}

\section{Introduction}
\label{sec:introduction}

Regulatory frameworks governing environmental, financial, and social conduct increasingly demand automated systems that can interpret complex legal texts and extract structured compliance requirements~\citep{zhong2020legal, katz2024gpt4law}.
The EU Taxonomy Regulation~\citep{eu2020taxonomy} exemplifies this challenge: its three Delegated Acts~\citep{eu2021climate, eu2022complementary, eu2023environmental} define technical screening criteria for hundreds of economic activities across six environmental objectives.
These criteria are expressed in complex natural language with nested structures combining quantitative thresholds, implicit hierarchical groupings, cross-references to external regulations, and evaluation logic that requires understanding logical relationships among sub-criteria ~(see Appendix~\ref{app:example} for an example).


Despite strong advances in legal reasoning~\citep{katz2024gpt4law, blairstanek2023statutory} and LLM-based extraction~\citep{josifoski2022genie, wadhwa2023revisiting}, applying language models to regulatory structuring exposes four key limitations:
(i)~\emph{structural hallucination}, where a single LLM call cannot reliably maintain hierarchical relationships across dozens of criteria;
(ii)~\emph{cross-stage inconsistency}, where early errors propagate uncorrected through later stages;
(iii)~\emph{structural violation}, where LLM reasoning alone cannot guarantee global properties such as hierarchy well-formedness or cycle-free dependencies; and (iv)~\emph{reference incompleteness}, where criteria reference external legal acts, internal sections, and footnotes, leaving the output dependent on dozens of additional documents. To the best of our knowledge, no existing system produces a \emph{self-contained} output that resolves all such dependencies inline.


We present \framework{} (\textbf{Reg}ulatory \textbf{Re}asoning with \textbf{A}gentic \textbf{C}orrec\textbf{t}ion), a multi-agent framework that addresses these challenges through three mechanisms: (1)~a seven-stage pipeline with specialized agents, each equipped with an \emph{Observe--Diagnose--Repair} (\odr{}) self-correction loop grounded in source document evidence, and a shared semantic memory for cross-stage consistency~(\S\ref{subsec:odr}, \S\ref{subsec:shared_context}); (2)~a typed criterion graph encoding six relationship types that enforces structural constraints such as hierarchy well-formedness and cycle-free inheritance~(\S\ref{subsec:graph}); and (3)~criterion-conditioned RAG that resolves every external reference, internal cross-reference, and footnote dependency inline, producing fully \emph{self-contained} outputs~(\S\ref{subsec:rag}).


Applying \framework{} to the three EU Taxonomy Delegated Acts, we construct \dataset{} dataset, covering all 242 activities with hierarchical criteria, thresholds, dependencies, and resolved source summaries. 
Table~\ref{tab:comparison} summarizes how \framework{} differs from prior systems across multi-stage decomposition, iterative correction, hierarchical nesting, cross-reference handling, and self-contained output.

\paragraph{Contributions.} 
 \textbf{ (1)}~A self-correcting multi-agent pipeline with source-grounded ODR loops~(\S\ref{subsec:odr}).
\textbf{(2)}~Criterion-conditioned RAG that produces self-contained outputs requiring no additional document consultation~(\S\ref{subsec:rag}).
\textbf{(3)}~\dataset{}, a structured dataset covering 242 EU Taxonomy activities with hierarchical criteria, thresholds, dependencies, and enriched source summaries~(\S\ref{sec:dataset}).

\section{Related Work}
\label{sec:related}

\framework{} lies at the intersection of legal and regulatory NLP, structured information extraction, self-correction, and retrieval-augmented generation. Table~\ref{tab:comparison} summarizes representative prior systems, with a more detailed discussion provided in Appendix ~\ref{app:related_detail}.

\begin{table*}[t]
\centering
\small
\setlength{\tabcolsep}{4.5pt}
\caption{Comparison of regulatory extraction systems on key problem dimensions.
\cmark{} = full support, $\sim$ = partial, \xmark{} = not addressed.
Reg.\ = primary regulatory text; Struct.\ = structured output; Decomp.\ =
multi-stage decomposition; Corr.\ = iterative correction; Hier.\ =
hierarchical nesting; Cross-ref.\ = cross-reference handling; GraphCompl.\ =
GraphCompliance \citep{chung2025graphcompliance}.}
\label{tab:comparison}
\begin{tabular}{@{}l ccccccc @{}}
\toprule
\multirow{2}{*}{\textbf{System}} & \multicolumn{2}{c}{\textit{Task scope}} & \multicolumn{2}{c}{\textit{Extraction approach}} & \multicolumn{3}{c}{\textit{Output properties}} \\
\cmidrule(lr){2-3}\cmidrule(lr){4-5}\cmidrule(lr){6-8}
 & \makecell{Reg.\\text} & \makecell{Struct.\\output} & \makecell{Multi-stage\\decomp.} & \makecell{Iterative\\correction} & \makecell{Hier.\\nesting} & \makecell{Cross-ref.\\handling} & \makecell{Self-\\contained} \\
\midrule
XTRAREG \citep{araujo2025xtrareg}       & \cmark & \cmark & \xmark & \xmark & \xmark & $\sim$ & \xmark \\
Galli et al.\ \citep{galli2025aiact}    & \cmark & \cmark & \cmark & \xmark & $\sim$ & $\sim$ & \xmark \\
GraphCompl.\ \citep{chung2025graphcompliance} & \cmark & \cmark & \cmark & \xmark & $\sim$ & \cmark & \xmark \\
AgenticIE \citep{colakoglu2025agenticie} & $\sim$ & \cmark & \cmark & \cmark & $\sim$ & \xmark & \xmark \\
P2T \citep{datla2025p2t}                & \cmark & \cmark & \cmark & \cmark & \xmark & \xmark & \xmark \\
L4M \citep{chen2025l4m}                 & $\sim$ & $\sim$ & \cmark & \cmark & \xmark & $\sim$ & \xmark \\
\midrule
\textbf{\framework{}} (ours)            & \cmark & \cmark & \cmark & \cmark & \cmark & \cmark & \cmark \\
\bottomrule
\end{tabular}
\end{table*}


\paragraph{Legal NLP and Regulatory Extraction.}
Despite advances from domain-specific pre-training~\citep{chalkidis2020legalbert}, multi-task benchmarks~\citep{chalkidis2022lexglue, guha2024legalbench}, and LLM-based reasoning~\citep{katz2024gpt4law}, most work targets passage-level tasks~\citep{hendrycks2021cuad}, leaving hierarchical compliance extraction largely unexplored.
Existing approaches each cover only part of the problem: XTRAREG~\citep{araujo2025xtrareg} and
\citet{galli2025aiact} extract regulatory requirements but produce flat or shallow structures without deep
hierarchical nesting; GraphCompliance~\citep{chung2025graphcompliance} uses graph structure for cross-references
but targets compliance checking rather than extraction; and AgenticIE~\citep{colakoglu2025agenticie} and
P2T~\citep{datla2025p2t} incorporate correction loops but lack deep hierarchical output (Appendix~\ref{app:related_detail}).
\citet{schmoll2025taxokpi} further show that LLMs achieve only moderate success on EU Taxonomy activity identification, motivating structured multi-agent approaches.


\paragraph{Self-Correction and RAG.} Self-Refine~\citep{madaan2023selfrefine} and Reflexion~\citep{shinn2023reflexion} established iterative refinement via self-generated feedback, but \citet{huang2024selfcorrect} showed that LLMs struggle to self-correct without external signals, confirmed by CRITIC~\citep{gou2024critic}. Our \odr{} mechanism extends this principle to regulatory extraction using source-document comparison and domain-specific issue taxonomies.
On the retrieval side, standard RAG~\citep{lewis2020rag, gao2024ragsurvey} retrieves context for generation and then discards it; adaptive strategies trigger re-retrieval when confidence drops~\citep{jiang2023flare, asai2024selfrag, yan2024crag}. Our criterion-conditioned RAG differs by treating each criterion as an independent retrieval query, producing criterion-focused summaries, and persisting the resolved content inline as a permanent part of the output.

\section{Methodology}
\label{sec:methodology}

\subsection{Problem Formulation}
\label{subsec:problem}

Let $\mathcal{D}$ be a regulatory document comprising extraction units $\mathcal{A} = \{a_1, \ldots, a_n\}$. An extraction unit may correspond to an activity, provision, obligation, or regulatory section depending on the document structure. In the EU Taxonomy setting considered here, each $a_i$ is an economic activity
represented as a semi-structured HTML fragment $h_i$.
Our objective is to extract, for each extraction unit, a structured representation $\mathcal{S}_i = (\mathcal{C}_i, \phi_i, \mathcal{T}_i, \mathcal{R}_i, \mathcal{G}_i, \xi_i)$ that encodes six components: hierarchical criteria $\mathcal{C}_i$ with evaluation logic $\phi_i$, quantitative and temporal thresholds $\mathcal{T}_i$, external references and inter-criteria dependencies $\mathcal{R}_i$, a typed criterion graph $\mathcal{G}_i$ encoding structural relationships, and criterion-focused summaries $\xi_i$ that resolve every reference inline.
This ensures the \emph{self-containedness property}: all information needed for compliance assessment is embedded within~$\mathcal{S}_i$.

This task is challenging for several reasons: \textbf{(i)}~implicit hierarchical nesting signaled by formatting rather than explicit markup; \textbf{(ii)}~\emph{unnumbered paragraphs} that must be semantically anchored to the correct parent; \textbf{(iii)}~thresholds stated in paragraphs distant from the criteria they constrain; \textbf{(iv)}~evaluation logic expressed implicitly through natural language connectives (``all of the following'' vs.\ ``one of the following pathways''); and \textbf{(v)}~heterogeneous citation formats requiring normalization to machine-readable identifiers.


\subsection{Framework Overview}
  \label{subsec:overview}

\begin{figure*}[t]
      \centering
      \includegraphics[width=\textwidth]{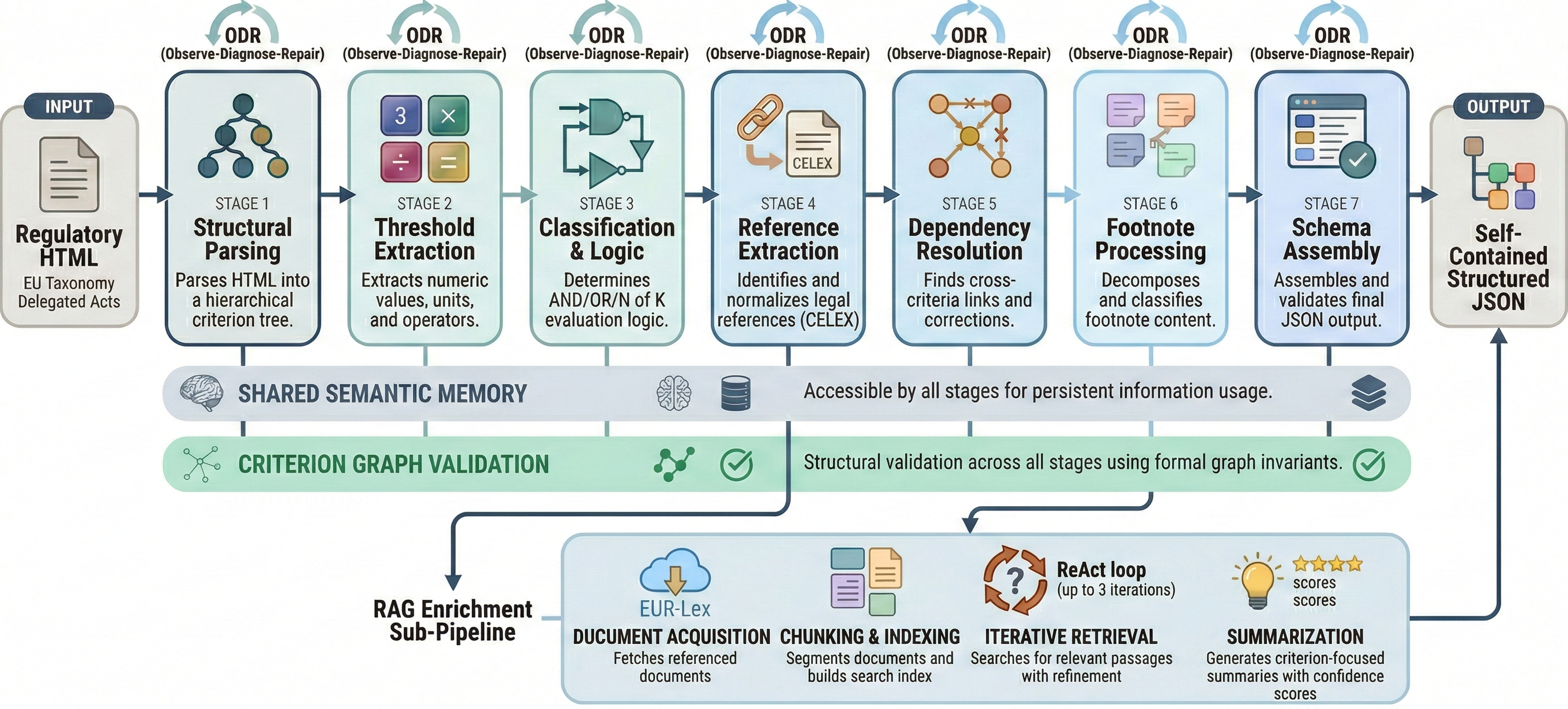}
      \caption{Overview of the \framework{} pipeline. Each stage is executed by a specialized agent with an \odr{} self-correction loop. The shared semantic memory maintains cross-stage consistency, the criterion graph enforces structural invariants, and the RAG sub-pipeline resolves external references inline to produce self-contained output.}
      \label{fig:overview}
  \end{figure*}

\framework{} addresses these challenges with a seven-stage pipeline (Figure~\ref{fig:overview}) organized around five principles for reliable structured extraction.
\textbf{Decomposition:} rather than prompting a single LLM to produce the output in one pass, we decompose extraction into stages aligned with distinct linguistic and structural competencies (\S\ref{subsec:agents}).
\textbf{Self-correction:} each stage implements an \odr{} loop (\S\ref{subsec:odr}) that detects and repairs errors by comparing output against source evidence.
\textbf{Structural enforcement:} a typed criterion graph
(\S\ref{subsec:graph}) catches structural violations that LLM reasoning alone cannot reliably detect.
\textbf{Self-containedness:} every external reference, cross-reference, and footnote dependency is resolved inline through criterion-conditioned RAG (\S\ref{subsec:rag}).
\textbf{Cross-stage consistency:}
\label{subsec:shared_context}
a \emph{shared semantic memory} maintains three registries (thresholds, cross-reference mappings, and activity metadata) populated incrementally and included in subsequent prompts.

\subsection{Specialized Agent Pipeline}
\label{subsec:agents}

The pipeline comprises seven stages---structural parsing, threshold extraction, content classification, reference extraction, dependency resolution, footnote processing, and schema assembly---designed to mirror the cognitive process a regulatory expert follows when manually structuring compliance criteria (see Table~\ref{tab:agents} in Appendix~\ref{app:method_details} for a summary).

\paragraph{Stage 1: Structural Parsing.}
The first agent parses the regulation's HTML and recovers the logical criterion
hierarchy. Although the markup contains ordered lists, its nesting does not
reliably reflect the actual structure, which is conveyed through typographic cues
such as indentation and numbering patterns. For instance, two mutually exclusive
compliance pathways may share a single \texttt{<ol>} element with no markup
distinguishing them, so the agent must infer parent--child relationships from
these cues, producing structured identifiers (e.g., \texttt{SC.1.a.ii}).


A further challenge is \emph{unnumbered paragraphs}: regulatory passages carrying substantive compliance content---verification obligations, methodology prescriptions, or background conditions---yet lacking any identifier linking them to the criteria they govern. Since identical phrasing can serve different functions depending on position and context, the agent uses semantic analysis to anchor each paragraph to the appropriate criterion, place it at the correct level in the hierarchy, assign it one of six descriptive types (Verification, Methodology, Commitment, Assessment, Replacement, or BackgroundInformation), and generate a stable identifier such as \texttt{1(f).Verification} (see Appendix~\ref{app:anchoring}).


\paragraph{Stage 2: Threshold Extraction.}
Regulatory criteria often embed quantitative limits and temporal deadlines in diverse surface forms (e.g., ``not exceeding 100 g CO$_2$e/kWh'', ``by 31 December 2030'') that may appear far from the criterion they constrain or apply to multiple criteria through inheritance. The agent extracts and normalizes them into structured representations, verifying each against the source text to reject hallucinated quantities, and registers results in the shared semantic memory~(\S\ref{subsec:shared_context}) for cross-stage consistency.

\paragraph{Stage 3: Classification and Logic Inference.}
Beyond identifying criteria, compliance assessment requires determining their types and how they combine. The agent classifies each criterion along two dimensions---\emph{category} (Quantitative vs.\ Qualitative) and \emph{applicability} (Mandatory vs.\ Conditional)---and annotates it with the \emph{evaluation logic} $\phi(c)$ that governs how its children combine.
Category assignment is grounded in Stage~2 output: a criterion is Quantitative iff it carries quantitative threshold data. Applicability is inferred from the criterion's text and hierarchical position: criteria scoped by conditional phrases (e.g., ``Where\ldots'', ``If\ldots'') are marked Conditional, while criteria that apply unconditionally are marked Mandatory.

The evaluation logic $\phi(c)$ (\textsc{and}, \textsc{or}, or \textsc{n\_of\_k}) determines whether all, any, or a specific count of a criterion's children must be satisfied.
This is particularly challenging when the text is misleading: a criterion may state ``the activity meets either of the following'' and list children (a) through (f), suggesting a simple choice among six options, when in reality (a--e) form one joint pathway and (f) defines a separate pathway with its own sub-criteria (see Appendix~\ref{app:example}). When such alternatives are detected, a \emph{semantic pathway detection} pass identifies which children belong to each
group, producing \textsc{group\_member} edges in the criterion graph (see Appendix~\ref{app:example}).

\paragraph{Stage 4: Reference Extraction.} Regulatory criteria frequently delegate requirements to external legal instruments (``in accordance with Directive 2010/75/EU''), international standards (``ISO 14064-1''), or internal cross-references (``Section 5.11 of this Annex''). Not all references carry equal weight: some define operative requirements the criterion depends on, while others serve only as background context. The agent classifies each reference accordingly as \texttt{must\_be\_fetched}---when the criterion cannot be assessed without consulting the source---or \texttt{citation\_only}---when it provides supplementary context only. This classification determines which references enter the retrieval pipeline~(\S\ref{subsec:rag}).

For EU legislation, references are normalized to machine-readable CELEX identifiers (e.g., ``Regulation (EU) 2018/1999'' $\to$ \texttt{32018R1999}) through multi-step parsing and validation (see Appendix~\ref{app:celex}). Each identifier is then verified against EUR-Lex to confirm the referenced document exists. If verification fails, the pipeline enters a ReAct-style correction loop: it observes the error, reasons about its likely cause---such as a regulation misclassified as a directive or a malformed year---applies a targeted fix, and re-verifies. This repeats for up to three attempts; if all fail, the pipeline falls back to a title-based EUR-Lex search. References that remain unresolved are flagged for manual review.



\paragraph{Stage 5: Dependency Resolution.}
Criteria in regulatory documents rarely stand alone: one criterion may apply only if another is satisfied, and some thresholds are defined in one clause yet silently referenced by another. Since these relationships are never explicitly marked in the HTML, the agent infers them from the text and organizes them into three types.
\emph{Conditional dependencies} arise when a criterion activates only under specific conditions---for example, a verification obligation that applies only when the transitional compliance pathway is selected (``if the facility complies via point~1(f)\ldots'').
\emph{Threshold inheritance} occurs when a criterion points to a value defined elsewhere (e.g., ``the emissions referred to in point~1(f)(a)''); the agent copies the referenced threshold and tags it with a \texttt{[THRESHOLD\_FROM:1(f)(a)]} annotation.
\emph{Cross-reference corrections} handle a less obvious problem: errors in the regulation itself, where a criterion is cited under the wrong identifier---for instance, ``point~1(b)'' when the actual target is criterion~1(f). The agent detects such mismatches through semantic analysis and records every correction with full provenance (e.g., \texttt{[CORR FROM:1(b) TO:1(f)]}, see Appendix~\ref{app:example}).
All three relationship types are encoded as typed edges in the criterion graph~(\S\ref{subsec:graph}).


\paragraph{Stage 6: Footnote Processing.}
Footnotes in regulatory documents often contain critical compliance details—legal references, technical definitions, official standards, and explanatory notes—embedded as hidden text within the HTML. A single footnote may combine multiple distinct items of different types: for instance, footnote~230 in the Appendix example contains 13 references mixing EU legal acts with international standards, each requiring independent classification as \texttt{must\_be\_fetched} or \texttt{citation\_only}. The agent extracts each footnote's content, decomposes it into individual items, categorizes them, and links each footnote to the criterion whose text contains its marker.


\paragraph{Stage 7: Schema Assembly.}
The final agent assembles all extracted components into a validated JSON schema, performing format normalization and cross-field consistency checks. These checks verify that category labels agree with threshold data, that evaluation logic matches child count, and that complex objects with no meaningful content are nullified.


\subsection{Self-Correction via ODR}
\label{subsec:odr}

Each agent's output is iteratively refined through a self-correction loop. Building on ReAct~\citep{yao2023react} and iterative self-refinement~\citep{madaan2023selfrefine, shinn2023reflexion}, \odr{} adapts these ideas to structured information extraction: rather than relying solely on the model's self-assessment, the agent compares its output against the source document to ground error detection in verifiable evidence.


\paragraph{Observe--Diagnose--Repair.}
In the \textbf{observe} step, the agent compares its structured output against the original HTML and enumerates discrepancies categorized by type and severity (see Appendix~\ref{app:odr_trace} for the full issue taxonomy).
This provides the external feedback signal that~\citet{huang2024selfcorrect} identify as necessary for effective self-correction.
In the \textbf{diagnose} step, the LLM analyzes observed issues and previous attempts to identify root causes (see Appendix~\ref{app:prompt_diagnose}).
In the \textbf{repair} step, the agent uses the diagnosis to construct a targeted correction prompt, and the agent re-extracts with this guidance.

The loop terminates when confidence $\geq \tau$ with no critical issues, when the same issues repeat, or upon timeout, returning the highest-confidence checkpoint.
When the loop exhausts its iterations without resolving critical issues, the output is flagged for human review rather than silently accepted.
We use $\tau = 0.7$ and $k_{\max} = 3$ throughout our experiments.

\subsection{Graph-Based Structural Validation}
\label{subsec:graph}

Since each agent operates independently, their outputs may contradict each
other. For example, the evaluation logic agent (Stage~3) may determine that a
criterion requires all three of its children to be satisfied, while the
structural parser (Stage~1) actually extracted four children for that criterion.
Similarly, a dependency identified in Stage~5 may point to a criterion that was
never extracted. To catch such inconsistencies, we unify all stage outputs into
a single criterion graph $\mathcal{G}_i = (V_i, E_i)$ per activity, where
vertices represent criteria and edges encode six relationship types:
\textsc{Hierarchy} (parent--child nesting),
\textsc{Group\_Member} (sibling grouping under shared evaluation logic),
\textsc{Inherits\_Threshold} (threshold propagation between criteria),
\textsc{References} (citation of external sources),
\textsc{Depends\_On} (conditional inter-criteria dependencies),
and \textsc{Corrects} (cross-reference corrections identified by semantic
analysis).

\paragraph{Construction and Validation.}

As each stage completes, its relationships are added as typed
edges---hierarchy from Stage~1, grouping from Stage~3, references from Stage~4, and dependencies, threshold inheritance and cross-reference corrections from Stage~5---with per-edge checks ensuring that the target criterion exists.
The graph also repositions criteria that the flat HTML caused to be mis-nested, using their identifiers to restore the correct hierarchy, and links unnumbered paragraphs to the correct parent. Once fully constructed, a global validation pass checks consistency at two levels: \emph{structurally}, verifying that hierarchy edges form a cycle-free tree, that the evaluation logic of each
criterion matches its actual child count, and that no criteria are left disconnected; and at the \emph{regulatory} level, enforcing task-specific constraints including an \emph{Evaluation Participation Rule} that excludes
ancillary nodes (e.g., background information) from child counts so that non-binding content does not distort the compliance logic inferred in Stage~3.

\subsection{Criterion-Conditioned RAG}
\label{subsec:rag}

Regulatory criteria frequently refer to external legal instruments
(e.g., ``in accordance with Directive 2010/75/EU'') and to other sections 
of the same regulation (e.g., ``Section 5.11 of this Annex'') without 
restating the referenced content. To make the output self-contained, 
the RAG module retrieves and embeds the relevant content directly 
within each criterion through a four-stage pipeline, operating on 
references from Stage~4 and footnote items from Stage~6 that carry 
CELEX identifiers.

\paragraph{Document Acquisition.}
Referenced documents are identified by their CELEX identifiers and 
fetched in PDF format from EUR-Lex, then converted to structured 
Markdown using MinerU~\citep{wang2024mineru}.

\paragraph{Semantic Chunking and Indexing.}
Documents are chunked using a structure-aware strategy that respects 
article and paragraph boundaries. Chunks are encoded with a ColBERT late-interaction model~\citep{khattab2020colbert} and indexed using Hierarchical Navigable Small World (HNSW) approximate nearest-neighbour search~\citep{malkov2020hnsw} on a per-document basis.

\paragraph{Iterative Criterion-Conditioned Retrieval.}
Rather than issuing a generic query, the LLM rewrites each criterion 
into a retrieval-optimized natural-language question enriched with 
the activity name, environmental objective, and article reference 
(e.g., \emph{``What emission threshold does Article~29(4)(a) 
establish for electricity generation using solar photovoltaic 
technology?''}). 
Each retrieval iteration fuses ColBERT dense scores with BM25 sparse scores via reciprocal rank fusion (RRF;~\citealt{cormack2009rrf}) and reranks candidates using ColBERT MaxSim scoring.
Retrieval follows an iterative refinement loop inspired by ReAct~\citep{yao2023react}: the LLM evaluates retrieved passages against the criterion and identifies information gaps (e.g., ``threshold value not found'', ``temporal constraint missing''). If confidence remains below a 
threshold, the gap description generates a refined query, and retrieval is repeated for up to three iterations.

\paragraph{Query-Focused Summarization.}
The top-ranked chunks are passed to the LLM together with the 
original criterion, so that the generated summary retains only the 
information relevant to that specific compliance requirement---
thresholds, definitions, exceptions, and conditions. To preserve 
factual accuracy, the summarizer enforces a verbatim quote limit, 
ensuring that summaries are genuine distillations rather than 
reproductions. The resolved content is persisted inline, eliminating 
the need to consult the original reference documents.

\section{The \dataset{} Dataset}
\label{sec:dataset}



We construct \dataset{} by applying \framework{} to the three EU Taxonomy Delegated Acts~\citep{eu2021climate, eu2022complementary, eu2023environmental}, a domain that concentrates all four challenges identified in \S\ref{sec:introduction}. The resulting dataset covers 242 activities with over 4,800 hierarchical criterion nodes across six environmental objectives, each encoding evaluation logic, thresholds, inter-criteria dependencies, and resolved reference summaries inline (output schema in Appendix~\ref{app:schema}).


\paragraph{Gold Annotation.}
To evaluate extraction quality, we constructed a gold-annotated subset of 100 activities via stratified sampling across three structural complexity tiers and all three Delegated Acts. Two PhD students served as annotators, each
independently reviewing 50 of the 100 activities across nine evaluation dimensions following the extraction specification (Appendix~\ref{app:schema}). Inter-annotator agreement, measured on a shared subset of 15 activities annotated by both, yielded Cohen's $\kappa = 0.84\text{--}0.91$ across three core dimensions (Appendix~\ref{app:annotation}).

\section{Experimental Setup}
\label{sec:experiments}

\paragraph{Evaluation Strategy.}
Direct comparison with related work (Table~\ref{tab:comparison}) is infeasible
because each operates on different regulatory texts with incompatible output schemas.
We instead evaluate on $n{=}100$ manually annotated activities and compare against a
\emph{GPT-4o single-pass} baseline that receives the same schema and prompt in one call,
testing whether a larger model alone can match multi-agent decomposition.
We evaluate along three axes: structural and classification accuracy, semantic equivalence via an LLM judge, and RAG summary quality.


\paragraph{Structural \& Classification Metrics ($n{=}100$).}
Four metrics are computed against gold annotations.
\textbf{Structural F1} aligns predicted and gold criterion trees by \texttt{criterion\_id} matching, scoring each pair in $[0,1]$ based on placement correctness and schema completeness (Appendix~\ref{app:structural_f1}).
\textbf{Category Accuracy} measures correct Quantitative/Qualitative assignment.
\textbf{Applicability Accuracy} measures correct Mandatory/Conditional assignment, which distinguishes universally applicable criteria from those gated by conditions.
\textbf{Evaluation Logic Accuracy} checks whether the predicted logic (\textsc{and}, \textsc{or}, \textsc{n\_of\_k}, \textsc{leaf}) matches gold, testing the ability to distinguish conjunctive from disjunctive requirements.


\paragraph{Semantic Equivalence Metrics ($n{\leq}100$).}
A GPT-4o judge scores four dimensions on a 0--5 scale (0 = both sides empty; 1--5 from wrong to full equivalence), receiving the source text, gold field, and extracted field (rubrics in Appendix~\ref{app:judge_prompts}). Scores are averaged per activity over non-zero nodes; $n$ varies across dimensions.
\textbf{Threshold} checks value, unit, operator, and temporal constraints.
\textbf{Reference} verifies completeness and correctness of legal citations, including CELEX normalization and \texttt{must\_be\_fetched}/\texttt{citation\_only} classification.
\textbf{Footnote} checks faithful capture and correct criterion linkage.
\textbf{Dependency} assesses conditional gating, threshold inheritance, and cross-reference corrections against the gold annotation.

\paragraph{RAG Summary Quality.}
Each generated summary is scored by a GPT-4o judge on a 1--5 Likert scale (prompts in Appendix~\ref{app:rag_prompts}) along four dimensions.
\textbf{Faithfulness} measures the proportion of summary claims grounded in retrieved
passages, penalizing hallucination but not omission.
\textbf{Relevance} classifies claims into three tiers (directly relevant, contextual,
unrelated) relative to the criterion.
\textbf{Completeness} evaluates whether compliance-critical information is captured,
weighted by element importance (major vs.\ minor).
\textbf{Coverage} measures whether the summary addresses the criterion's information
needs, weighted by need priority (primary vs.\ secondary).

\section{Results and Analysis}
\label{sec:results}

\subsection{Extraction Quality}
\label{subsec:extraction}

\begin{table}[t]
\centering
\caption{Evaluation results for \framework{} on $n{=}100$ gold-annotated activities. Structural and classification metrics are percentages; semantic equivalence uses a 1--5 GPT-4o judge scale (\S5). Activities with all-zero scores are excluded, so $n$ varies. }
\label{tab:main_results}
\resizebox{\columnwidth}{!}{%
\begin{tabular}{@{}lcccc@{}}
\toprule
\textbf{Metric} & \textbf{Mean} & \textbf{Std} & \textbf{Median} & \textbf{$n$} \\
\midrule
\rowcolor{gray!12}\multicolumn{5}{c}{\textit{Structural \& Classification (\%)}} \\
Structural F1 & 94.12 & 3.56 & 95.2 & 100 \\
Category Accuracy & 98.6 & 2.62 & 100.0 & 100 \\
Applicability Accuracy & 97.24 & 3.16 & 98.4 & 100 \\
Evaluation Logic Accuracy & 93.4 & 3.60 & 94.0 & 100 \\
\midrule
\rowcolor{gray!12}\multicolumn{5}{c}{\textit{Semantic Equivalence (1--5)}} \\
Threshold & 4.43 & 0.54 & 5.00 & 74 \\
Reference & 4.77 & 0.22 & 5.00 & 100 \\
Footnote & 4.48 & 0.60 & 4.50 & 99 \\
Dependency & 4.63 & 0.80 & 5.00 & 99 \\
\bottomrule
\end{tabular}}
\end{table}

\paragraph{Structural \& Classification Accuracy.}
On the 100 gold-annotated activities (Table~\ref{tab:main_results}), \framework{} achieves 94.12\% \textit{structural F1}, reconstructing criterion trees with high fidelity; residual errors concentrate in activities with deeply nested alternative pathways.
\textit{Category accuracy} is highest at 98.6\%, as the deterministic category rule in Stage~3 leaves little room for error once thresholds have been correctly extracted.
\textit{Applicability accuracy} reaches 97.24\%; residual errors arise when conditional language is ambiguous between a compliance requirement and an applicability condition.
\textit{Evaluation logic accuracy} (93.4\%) reflects the difficulty of distinguishing conjunctive from disjunctive requirements in ambiguous (chapeau) text.

\paragraph{Semantic Equivalence.}
The LLM-judge scores confirm strong semantic fidelity across all four dimensions, with all means exceeding 4.4 on the 1--5 scale.
Reference equivalence achieves 4.77, the highest among the four dimensions, reflecting the effectiveness of deterministic CELEX normalization.
Dependency equivalence reaches 4.63, reflecting the inherent complexity of dependency chains spanning multiple criteria.
Threshold equivalence scores 4.43 ($n{=}74$), with the multi-component nature of thresholds (value, unit, operator, temporal constraint) making this the most demanding semantic dimension.
Footnote equivalence is 4.48, with residual errors arising from footnotes that combine multiple items of different types.

\paragraph{Baseline Comparison.}

\begin{table}[t]
\centering
\caption{\framework{}  vs. GPT-4o single-pass baseline ($n{=}100$ gold-annotated activities). Structural and classification metrics are percentages; semantic equivalence as mean Likert scores (1--5).}
\label{tab:baseline}
\resizebox{\columnwidth}{!}{%
\begin{tabular}{@{}lccc@{}}
\toprule
\textbf{Metric} & \textbf{\framework{}} & \textbf{GPT-4o} & \textbf{$\Delta$} \\
\midrule
\rowcolor{gray!12}\multicolumn{4}{c}{\textit{Structural \& Classification (\%)}} \\
Structural F1 & 94.12 & 78.6 & +15.52 \\
Category Accuracy & 98.6 & 90.2 & +8.4 \\
Applicability Accuracy & 97.24 & 85.7 & +11.54 \\
Evaluation Logic Accuracy & 93.4 & 80.3 & +13.1 \\
\midrule
\rowcolor{gray!12}\multicolumn{4}{c}{\textit{Semantic Equivalence (1--5)}} \\
Threshold & 4.43 & 3.23 & +1.20 \\
Reference & 4.77 & 3.34 & +1.43 \\
Footnote & 4.48 & 3.12 & +1.36 \\
Dependency & 4.63 & 2.96 & +1.67 \\
\bottomrule
\end{tabular}}
\end{table}

Table~\ref{tab:baseline} compares \framework{} against a GPT-4o single-pass baseline that receives identical instructions and output schema but produces everything in one call.
Although the baseline relies on a larger model, \framework{}---using DeepSeek-R1-Distill-Qwen-32B~\citep{deepseek2025r1}---outperforms it on all eight metrics, implying that decomposition, self-correction, and graph validation matter more than model scale for this task.

The largest structural gaps appear on structural F1 (+15.52) and evaluation logic (+13.1), both of which demand hierarchical consistency across dozens of criteria---precisely where a single call loses track of parent--child relationships.
On the semantic side, dependency equivalence gains the most (+1.67), as detecting conditional gating and threshold inheritance requires the dedicated dependency stage and shared memory absent from the baseline. The remaining dimensions each improve by over one point, reflecting the value of specialized agents for compound thresholds, multi-item references, and footnotes.
\subsection{RAG Enrichment Quality}
\label{subsec:rag_eval}

We evaluate the RAG enrichment module on 709 criterion--source pairs spanning 242 activities and 93 EUR-Lex documents.

\begin{table}[t]
\centering
\caption{RAG summary quality scores (GPT-4o judge, 1--5 Likert).}
\label{tab:rag_quality}
\resizebox{0.7\columnwidth}{!}{%
\begin{tabular}{@{}lccc@{}}
\toprule
\textbf{Metric} & \textbf{Mean} & \textbf{Std} & \textbf{Median} \\
\midrule
Faithfulness & 4.61 & 1.13 & 5 \\
Relevance & 4.14 & 1.02 & 4 \\
Completeness & 4.07 & 0.85 & 4 \\
Coverage & 4.01 & 1.08 & 4 \\
\bottomrule
\end{tabular}}
\end{table}


Faithfulness is the strongest dimension (4.61), indicating that summary claims are directly traceable to retrieved passages---critical in the regulatory domain, where hallucinated obligations could lead to incorrect compliance assessments.
Relevance and completeness both exceed 4.0, confirming that summaries are targeted and capture major compliance-critical elements, while coverage confirms that retrieved passages address each criterion's core information needs.

\paragraph{ReAct Loop Analysis.}
The retrieval pipeline employs a ReAct loop~\citep{yao2023react} that reformulates queries when confidence falls below a threshold. Table~\ref{tab:react} stratifies quality by iteration count.

\begin{table}[t]
\centering
\caption{Quality by ReAct iteration count. Ret.\ Conf.\ is the pipeline's self-assessed retrieval confidence.}
\label{tab:react}
\resizebox{\columnwidth}{!}{%
\begin{tabular}{@{}ccccccc@{}}
\toprule
\textbf{Iter.} & \textbf{Count (\%)} & \textbf{Faith.} & \textbf{Relev.} & \textbf{Compl.} & \textbf{Cover.} & \textbf{Ret.\ Conf.} \\
\midrule
1 & 535 (75.5\%) & 4.74 & 4.30 & 4.17 & 4.30 & 0.92 \\
2 & 88 (12.4\%) & 4.18 & 3.80 & 3.94 & 3.33 & 0.86 \\
3 & 86 (12.1\%) & 4.28 & 3.48 & 3.64 & 2.93 & 0.67 \\
\bottomrule
\end{tabular}}
\end{table}

Three observations emerge.
First, 75.5\% of pairs resolve in a single iteration with the highest quality across all metrics, indicating effective initial query formulation.
Second, multi-iteration cases appear to be harder retrieval tasks: relevance, completeness, and coverage decrease with iteration count, while the monotonic confidence decrease (0.92~$\to$~0.67) confirms that the pipeline correctly identifies difficult queries.
Third, faithfulness remains stable ($\geq 4.18$) across all iteration counts, showing that the pipeline maintains grounding quality under difficult retrieval rather than compensating with hallucination.

\begin{figure}[t]
    \centering
    \includegraphics[width=.8\columnwidth]{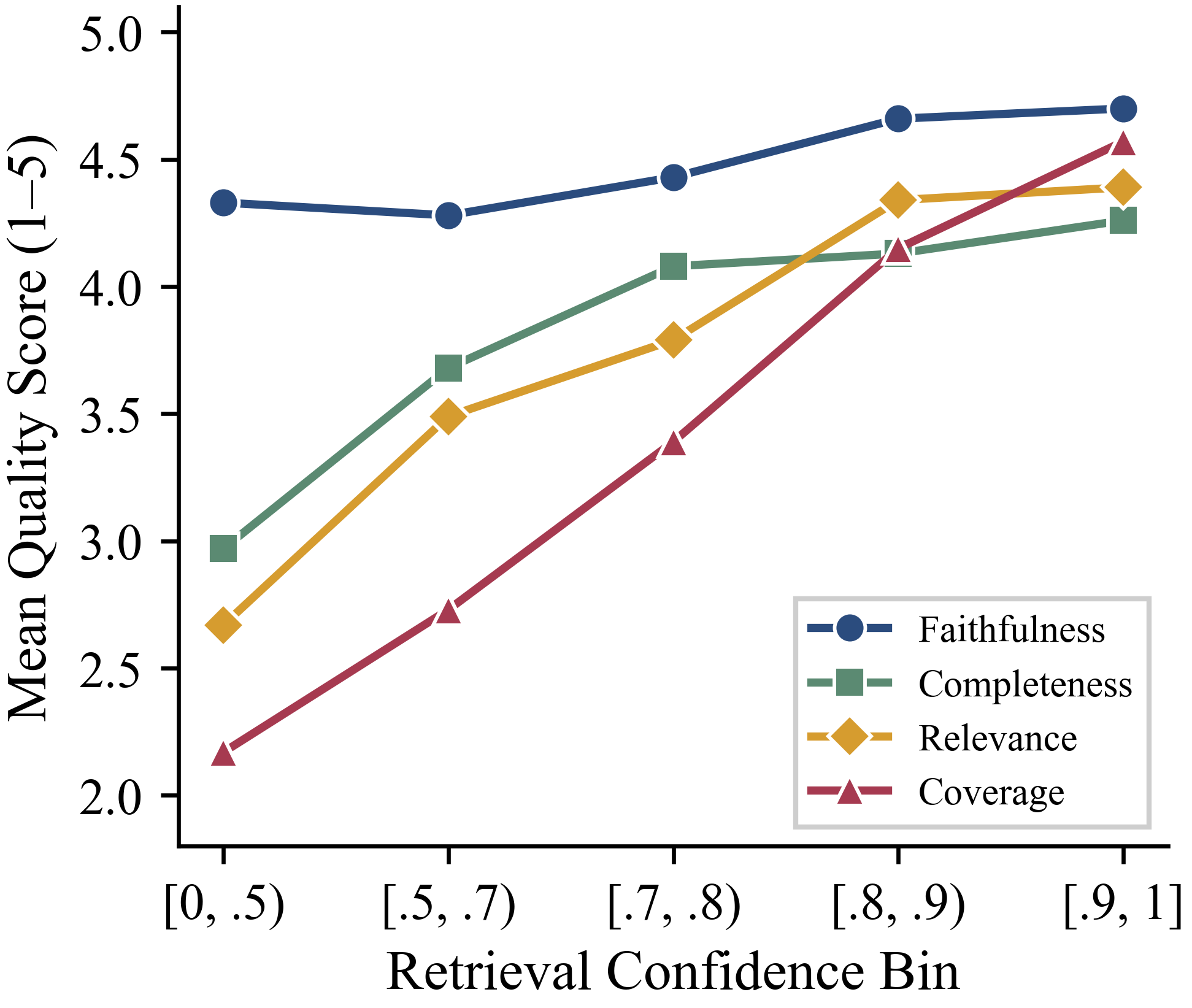}
    \caption{Mean quality scores per retrieval confidence bin, averaged over criterion–source pairs. Faithfulness remains stable while relevance, completeness, and coverage increase monotonically.}
   
    \label{fig:calibration}
\end{figure}

\paragraph{Confidence Calibration.}
\label{subsec:calibration}
We examine whether confidence scores predict output quality, enabling automated quality filtering.
Figure~\ref{fig:calibration} partitions the 709 criterion--source pairs into five confidence bins and reports the mean quality score per metric.

Relevance, completeness, and coverage increase monotonically across all confidence bins.
Faithfulness remains high ($\ge 4.28$) across all bins, varying less across confidence bins than the other three dimensions.
The clear separation between low- and high-confidence outputs, particularly for coverage (2.17 vs.\ 4.57), suggests that the confidence score provides a useful quality signal.

\section{Conclusion}


We presented \textsc{RegReAct}, a self-correcting multi-agent framework that decomposes regulatory information extraction into seven specialized stages with Observe--Diagnose--Repair loops grounded in source evidence, a typed criterion graph for structural validation, and criterion-conditioned RAG for inline reference resolution, producing self-contained structured outputs. Applying \textsc{RegReAct} to three EU Taxonomy Delegated Acts yielded \textsc{EU-TaxoStruct}, a dataset of 242 activities with hierarchical criteria, thresholds, and enriched source summaries. Evaluation against a GPT-4o single-pass baseline showed consistent gains across all metrics, highlighting the value of structured multi-agent pipelines over monolithic generation for complex regulatory texts. Future work will target other regulatory domains and languages.

\section*{Limitations}

Four practical limitations remain:
(1)~references to ISO/EN standards cannot be enriched inline as their content is behind paywalls, limiting self-containedness for those references; (2)~the current evaluation covers English-language regulations; multilingual EU texts remain untested; (3) as our evaluation is limited to the EU Taxonomy, the extent to which REGREACT generalizes to other regulatory domains remains to be established; and (4)~the extracted dataset reflects the regulations as published at the time of processing; subsequent amendments require re-execution of the pipeline.

\section*{Acknowledgments}
The authors would like to acknowledge the financial support provided by the Austrian Research Agency (FFG) for the project “AI Enabled Sustainability Jurisdiction Demonstrator” (project No. 915229). The computational results presented in this work have been achieved using the MUSICA cluster, part of the Austrian Scientific Computing (ASC) infrastructure.




\bibliography{references}

\clearpage
\appendix
\onecolumn

\section*{Appendix Contents}
\label{app:toc}
\apxline[app:related_detail]{}{0em}{2em}{A}{Extended Related Work}{\pageref{app:related_detail}}
\apxline[app:example]{}{0em}{2em}{B}{Example Extraction Output}{\pageref{app:example}}
\apxline[app:odr_trace]{}{0em}{2em}{C}{ODR Self-Correction Details}{\pageref{app:odr_trace}}
\apxline[app:prompts]{}{0em}{2em}{D}{Prompt Templates}{\pageref{app:prompts}}
\apxline[app:prompt_sse]{}{1em}{0em}{}{D.1--D.12~~Pipeline Stage Prompts (Stages 2--7)}{\pageref{app:prompt_sse}}
\apxline[app:prompt_observe]{}{1em}{0em}{}{D.13--D.14~~ODR Prompts}{\pageref{app:prompt_observe}}
\apxline[app:prompt_rewrite]{}{1em}{0em}{}{D.15--D.18~~RAG Prompts}{\pageref{app:prompt_rewrite}}
\apxline[app:judge_prompts]{}{1em}{0em}{}{D.19--D.20~~Evaluation Judge Prompts}{\pageref{app:judge_prompts}}
\apxline[app:method_details]{}{0em}{2em}{E}{Supplementary Methodology Details}{\pageref{app:method_details}}
\apxline[app:anchoring]{}{1em}{0em}{}{E.1~~Semantic Anchoring Procedure}{\pageref{app:anchoring}}
\apxline[app:celex]{}{1em}{0em}{}{E.2~~CELEX Identifier Parsing}{\pageref{app:celex}}
\apxline[app:structural_f1]{}{1em}{0em}{}{E.3~~Structural F1 Computation}{\pageref{app:structural_f1}}
\apxline[app:rag_eval]{}{1em}{0em}{}{E.4~~RAG Query Refinement}{\pageref{app:rag_eval}}
\apxline[app:schema]{}{0em}{2em}{F}{Output Schema}{\pageref{app:schema}}
\apxline[app:annotation]{}{0em}{2em}{G}{Gold Annotation Details}{\pageref{app:annotation}}


\twocolumn
\section{Extended Related Work}
\label{app:related_detail}

This appendix provides a detailed discussion of prior work organized by the four research threads that \framework{} builds upon.

\subsection{Legal and Regulatory NLP}

The application of NLP to the legal domain has progressed from rule-based systems to neural approaches.
LEGAL-BERT~\citep{chalkidis2020legalbert} demonstrated that domain-specific pre-training improves legal text classification, while LexGLUE~\citep{chalkidis2022lexglue} and LEXTREME~\citep{niklaus2023lextreme} established multi-task benchmarks across jurisdictions.
LegalBench~\citep{guha2024legalbench} expanded evaluation to 162 tasks, and CUAD~\citep{hendrycks2021cuad} provided expert-annotated data for contract clause identification.
At the LLM level, \citet{katz2024gpt4law} showed GPT-4 passing the Bar Examination, while \citet{blairstanek2023statutory} investigated statutory reasoning limitations.

The vast majority of this work operates at the \emph{classification} or \emph{question-answering} level over individual passages~\citep{zhong2020legal}.
The problem we address, extracting complete hierarchical compliance structures from regulation-length documents, remains largely unexplored.
Early work on automated legal metadata extraction~\citep{sleimi2018automated} targeted deontic statements but did not produce machine-readable compliance hierarchies.
XTRAREG~\citep{araujo2025xtrareg} used LLM+RAG to extract privacy requirements from the GDPR, achieving 82\% accuracy on access rights, but produces flat requirement lists without hierarchical nesting, inter-criteria dependencies, or self-contained output.
In the sustainability domain, \citet{schmoll2025taxokpi} evaluate LLMs on EU Taxonomy KPI extraction from 190 corporate reports, finding that models achieve only moderate success and function best as assistive tools. Complementing this, \citet{ali2025sustainableqa} introduce SustainableQA, a question answering benchmark over ESG and EU Taxonomy documents, and \citet{ali2025recor} present RECOR, a reasoning-intensive multi-turn conversational retrieval benchmark spanning sustainability among other domains; both works underscore the difficulty of regulatory NLP and motivate structured, reasoning-aware pipelines such as \framework{}.

\subsection{LLM-Based Regulatory Extraction}

Recent work has applied LLMs directly to regulatory information extraction. We organize the most relevant systems by their core contribution.

\paragraph{Multi-stage pipelines for EU legislation.}
\citet{galli2025aiact} present the most comprehensive pipeline for EU legislation, targeting the AI Act with a four-stage workflow: (1)~identification of obligation-bearing provisions, (2)~filtering of deontic statements, (3)~analysis of deontic content (addressees, predicates, conditions), and (4)~construction of a searchable knowledge graph.
Using LLaMA~3.3~70B with five expert evaluators, they achieve 93\% precision on obligation filtering and over 99\% accuracy on obligation type classification, identifying 729 obligations across the AI Act.
The knowledge graph captures article-to-obligation relationships, providing partial hierarchical structure;
however, obligations themselves are flat triples without nested sub-criteria, and the pipeline includes no
self-correction mechanism.

\paragraph{Graph-based compliance representations.}
GraphCompliance~\citep{chung2025graphcompliance} constructs a Policy Graph from GDPR provisions, encoding normative structure through typed nodes and cross-reference edges that link articles to exceptions, definitions, and related provisions.
A separate Context Graph formalizes runtime events as subject/action/object triples, and a deterministic Compliance Gate performs structural analysis via reference traversal and exception chaining.
GraphCompliance is the closest prior work to ours in its use of graph structure for cross-reference handling; however, it targets compliance \emph{checking} (producing compliant/non-compliant judgments) rather than criteria \emph{extraction}, and does not implement iterative correction or self-contained output.

\paragraph{Agentic self-correction on regulatory documents.}
AgenticIE~\citep{colakoglu2025agenticie} introduces a planner/executor/responder agent architecture for extracting key-value pairs from EU Declarations of Performance, standardized product certification forms issued under the Construction Products Regulation.
When the executor fails to extract a field, a self-repair loop re-prompts with failure feedback.
AgenticIE is the closest work to ours for agentic correction on EU regulatory documents; however, it targets
product certification forms rather than legislative text, uses a single adaptive agent rather than multiple
specialized agents, and produces shallow key-value structures (up to two levels of nesting for grouped product
properties) without deep hierarchical criterion trees or cross-reference resolution.

P2T~\citep{datla2025p2t} converts AI governance policies (EU AI Act, NIST AI RMF) into executable rules through a six-stage pipeline.
A mine/judge/repair loop iteratively refines extracted rules: an LLM judge scores each rule, and a repair agent addresses identified deficiencies.
An SMT solver then detects logical conflicts between rules.
P2T shares our philosophy of iterative correction with explicit quality assessment, but produces flat atomic rules without hierarchical nesting, cross-reference resolution, or self-contained output.

\paragraph{Legal agentic workflows.}
LAW~\citep{watson2024law} introduces agentic workflows for custody and fund services contracts, orchestrating specialized agents and domain-specific tools for structured extraction.
LAW demonstrates the effectiveness of agent specialization for legal extraction, but addresses a fundamentally different task (contract analysis with known clause types) and does not implement self-correction or graph-based validation.

General-purpose multi-agent frameworks~\citep{hong2024metagpt, wu2023autogen, chen2024agentverse} provide agent orchestration infrastructure but do not address structured output validation or domain-specific invariant enforcement required for regulatory extraction.

\subsection{Self-Correction in Language Models}

Self-Refine~\citep{madaan2023selfrefine} established iterative self-correction through self-generated feedback, while Reflexion~\citep{shinn2023reflexion} introduced verbal reinforcement learning to learn from prior failures.
ReAct~\citep{yao2023react} interleaves reasoning traces with actions, enabling models to ground outputs in external observations.
Chain-of-Thought~\citep{wei2022chain}, Self-Consistency~\citep{wang2023selfconsistency}, and Tree of Thoughts~\citep{yao2024tree} improve reasoning through deliberation and search.
\citet{welleck2023generating} formalized self-correction as a learnable generation strategy.

A critical finding by~\citet{huang2024selfcorrect} demonstrated that LLMs struggle to self-correct reasoning without external feedback signals.
\citet{pan2024correcting} survey the landscape, categorizing correction strategies by when they occur
(training-time, generation-time, and post-hoc) and whether feedback is self-generated or externally sourced.
CRITIC~\citep{gou2024critic} operationalized this insight by introducing a verify$\rightarrow$correct$\rightarrow$verify cycle that grounds correction in external tool interaction (search engines, code interpreters), achieving consistent improvements across QA, math, and toxicity reduction at ICLR 2024.

In the legal domain, L4M~\citep{chen2025l4m} is the closest system to ours in combining multi-agent architecture with iterative correction.
L4M embeds prosecutor- and defense-aligned LLM agents inside a logic-based control loop for case adjudication: agents extract facts and statutes into typed schemas, an SMT solver verifies soundness, and unsatisfiable cores trigger iterative self-critique until a consistent solution is reached.
L4M surpasses GPT-4o on Chinese criminal law judgment prediction through this combination of symbolic verification and neural reasoning.
However, L4M targets adjudication (determining case outcomes) rather than information extraction, and its correction mechanism relies on formal logic solvers that are not applicable to the structural and semantic errors that arise in regulatory extraction (missing criteria, incorrect hierarchical nesting, unresolved references).

Notably, among the extraction-focused systems reviewed above, only AgenticIE and P2T incorporate correction loops.
AgenticIE re-prompts failed extractions with failure feedback, and P2T applies an LLM judge to score and repair individual rules.
However, neither grounds its correction in the original source document: AgenticIE's repair is triggered by tool-execution failures rather than source comparison, and P2T's judge evaluates rule quality without cross-checking against the regulation text.
Our \odr{} mechanism extends the CRITIC principle to structured regulatory extraction.
Like CRITIC, \odr{} grounds correction in external evidence (the source HTML document); like L4M, it uses iterative refinement driven by external verification.
Unlike both, \odr{} operates within a multi-stage extraction pipeline where each agent has a domain-specific issue taxonomy (structural, semantic, completeness, consistency), a structured diagnosis step that identifies root causes and recommends actions, and a checkpoint mechanism that preserves the best intermediate result across iterations.

\subsection{Retrieval-Augmented Generation}

RAG~\citep{lewis2020rag} addresses LLM knowledge limitations by grounding generation in retrieved documents.
Dense passage retrieval~\citep{karpukhin2020dpr} and sentence embeddings~\citep{reimers2019sentencebert} enable semantic search, while~\citet{gao2024ragsurvey} catalogue advances across retrieval, augmentation, and generation strategies.
Recent work has moved toward \emph{adaptive} retrieval: FLARE~\citep{jiang2023flare} triggers retrieval when the generator's token-level confidence drops below a threshold, Self-RAG~\citep{asai2024selfrag} trains the model itself to decide when retrieval is needed via learned self-reflection tokens, and CRAG~\citep{yan2024crag} introduces a corrective retrieval mechanism that evaluates retrieval quality and triggers refined searches when initial results are insufficient.
In the legal domain, RAG has been applied to question answering over case law and legislation, though primarily for single-query retrieval rather than structured extraction.

Our criterion-conditioned RAG differs from standard approaches in three key respects.
First, each criterion serves as an independent retrieval query, fusing late-interaction token-level matching~\citep{khattab2020colbert} with sparse lexical signals via reciprocal rank fusion.
Second, retrieval is followed by \emph{criterion-focused summarization} that extracts only the thresholds, definitions, and conditions relevant to that specific compliance requirement, rather than returning generic document summaries.
Third, and most distinctively, the resolved content is \emph{persisted inline} within each criterion's structured output, producing self-contained representations.
While standard RAG retrieves context to assist generation and then discards it, our pipeline embeds the resolved content as a permanent component of the output, so downstream consumers require no access to the original reference documents.

\subsection{Markup-Aware Models and Graph-Based Retrieval}
\label{app:markup_graphrag}

Markup-aware language models such as MarkupLM~\citep{li2022markuplm} and WebFormer~\citep{wang2022webformer} jointly encode DOM structure and text to recover semantic roles in visually rich documents. 
These methods target settings where the markup carries the missing semantic signal. EUR-Lex HTML is structurally clean, but its markup does not encode the underlying logical hierarchy: as shown in Section~\ref{subsec:agents} (Stage~1) and Appendix~\ref{app:example}, two mutually exclusive compliance pathways routinely share a single \texttt{<ol>} with no markup cue between them. The parsing challenge is therefore semantic rather than markup-driven, which motivates the typographic and semantic anchoring used in Stage~1 instead of a markup-aware backbone.

Graph-based retrieval, exemplified by GraphRAG~\citep{edge2024graphrag}, 
retrieves over typed knowledge-graph edges rather than embedding similarity alone. Stages~5 and~6 of \framework{} follow this principle by constraining dependency resolution and footnote attribution to the six typed edges of the criterion graph (Section~\ref{subsec:graph}). 
The graph is built per-activity from the regulatory text, so the typed structure constrains retrieval during construction rather than against a pre-existing knowledge base.

\section{Example: End-to-End Extraction}
\label{app:example}

This appendix presents a complete input--output pair for Activity CCM~4.29 (\emph{Electricity generation from fossil gaseous fuels}), one of the most structurally complex activities in the EU Taxonomy.
The input (Figure~\ref{fig:input_regulation}) shows the regulation text as rendered for human readers; the output (Figures~\ref{fig:json_output_1}--\ref{lst:json_output}, spanning four pages) shows the structured JSON produced by \framework{}.

Key phenomena demonstrated:
\begin{itemize}[nosep,leftmargin=1.5em]
    \item \textbf{Semantic pathway detection}: criterion~1 contains two mutually exclusive pathways---\texttt{1(a--e)} (low-emission) vs.\ \texttt{1(f)} (transitional)---that share a single \texttt{<ol>} in the HTML but require an OR container with two AND groups.
    \item \textbf{Cross-reference correction}: the regulation text says ``point~1(b)'' in multiple places, but semantic analysis reveals these refer to \texttt{1(f)}; the output records \texttt{[CORR FROM:1(b) TO:1(f)]}.
    \item \textbf{Threshold inheritance}: verification sub-criterion \texttt{1(f).Verification(a)} inherits quantitative thresholds from \texttt{1(f)(a)} via \texttt{[THRESHOLD\_FROM:1(f)(a)]}.
    \item \textbf{Unnumbered paragraph anchoring}: two paragraphs between criteria~1 and~2 have no numbering in the source; they are anchored as \texttt{1(f).Verification} and \texttt{1(f).BackgroundInformation} with semantic type tags.
    \item \textbf{Dependencies}: ancillary nodes carry explicit dependency clauses (e.g., \texttt{1(f).Verification} depends on pathway \texttt{1(f)} being chosen).
    \item \textbf{OR-linked thresholds with periods}: criterion \texttt{1(f)(a)} has two alternative quantitative limits connected by OR logic, one with a 20-year bounded period.
    \item \textbf{Footnote structuring}: footnote~230 on criterion \texttt{1(f)(g)} is decomposed into 13 structured items distinguishing \texttt{must\_be\_fetched} from \texttt{citation\_only} references.
    \item \textbf{External references}: multiple criteria link to EU legislation with normalized CELEX identifiers and \texttt{must\_be\_fetched} / \texttt{citation\_only} typing.
\end{itemize}

\onecolumn
\begin{figure}[p]
\centering
\begin{tcolorbox}[
    title={\textbf{EU Taxonomy --- Activity CCM 4.29: Electricity generation from fossil gaseous fuels}\\[-0.3em]{\scriptsize Substantial contribution criteria (source: Delegated Regulation (EU) 2022/1214, Annex I)}},
    colback=white,colframe=black!70,fonttitle=\small,
    boxrule=0.4pt,arc=2pt,left=4pt,right=4pt,top=2pt,bottom=2pt,
    width=\textwidth
]
\footnotesize

\textbf{1.}~The activity meets either of the following criteria:

\begin{enumerate}[label=(\alph*),nosep,leftmargin=2em,itemsep=2pt]
    \item the life-cycle GHG emissions from the generation of electricity using fossil gaseous fuels are lower than 100\,g\,CO\textsubscript{2}e/kWh.
    \item Life-cycle GHG emissions are calculated based on project-specific data, where available, using Recommendation 2013/179/EU or, alternatively, using ISO~14067:2018 or ISO~14064-1:2018.
    \item Quantified life-cycle GHG emissions are verified by an independent third party.
    \item Where facilities incorporate any form of abatement, including carbon capture or use of renewable or low-carbon gases, that abatement activity complies with the criteria set out in the relevant Section of this Annex, where applicable.
    \item Where the CO\textsubscript{2} that would otherwise be emitted from the electricity generation process is captured for the purpose of underground storage, the CO\textsubscript{2} is transported and stored underground, in accordance with the technical screening criteria set out in Sections~5.11 and~5.12 of this Annex.
    \item facilities for which the construction permit is granted by 31~December~2030 comply with all of the following:
\end{enumerate}

\begin{enumerate}[label=(\alph*),nosep,leftmargin=2em,itemsep=2pt]
    \item direct GHG emissions of the activity are lower than 270\,g\,CO\textsubscript{2}e/kWh of the output energy, or annual direct GHG emissions of the activity do not exceed an average of 550\,kg\,CO\textsubscript{2}e/kW of the facility's capacity over 20~years;
    \item the power to be replaced cannot be generated from renewable energy sources, based on a comparative assessment with the most cost-effective and technically feasible renewable alternative for the same capacity identified; the result of this comparative assessment is published and is subject to a stakeholder consultation;
    \item the activity replaces an existing high emitting electricity generation activity that uses solid or liquid fossil fuels;
    \item the newly installed production capacity does not exceed the capacity of the replaced facility by more than 15\,\%;
    \item the facility is designed and constructed to use renewable and/or low-carbon gaseous fuels and the switch to full use of renewable and/or low-carbon gaseous fuels takes place by 31~December~2035, with a commitment and verifiable plan approved by the management body of the undertaking;
    \item the replacement leads to a reduction in emissions of at least 55\,\% GHG over the lifetime of the newly installed production capacity;
    \item where the activity takes place on the territory of a Member State in which coal is used for energy generation, that Member State has committed to phase-out the use of energy generation from coal and has reported this in its integrated national energy and climate plan referred to in Article~3 of Regulation~(EU)~2018/1999 of the European Parliament and of the Council\textsuperscript{(230)} or in another instrument.
\end{enumerate}

\smallskip
\noindent Compliance with the criteria referred to in point~1(b) is verified by an independent third party. The independent third-party verifier has the necessary resources and expertise to perform such verification. The independent third party verifier does not have any conflict of interest with the owner or the funder, and is not involved in the development or operation of the activity. The independent third party verifier carries out diligently the verification of compliance with the technical screening criteria. In particular, every year the independent third party publishes and transmits to the Commission a report:

\begin{enumerate}[label=(\alph*),nosep,leftmargin=2em,itemsep=2pt]
    \item certifying the level of direct GHG emissions referred to in point~1(b)(i);
    \item where applicable, assessing whether annual direct GHG emissions of the activity are on a credible trajectory to comply with the average threshold over 20~years referred to in point~1(b)(i);
    \item assessing whether the activity is on a credible trajectory to comply with point~1(b)(v).
\end{enumerate}

\smallskip
\noindent When undertaking the assessment referred to in point~1(b), the independent third party verifier takes into account in particular the planned annual direct GHG emissions for each year of the trajectory, realised annual direct GHG emissions, planned and realised operating hours, and planned and realised use of renewable or low carbon gases.

\smallskip
\noindent On the basis of the reports transmitted to it, the Commission may address an opinion to the relevant operators. The Commission shall take those reports into account when performing the review referred to in Article~19(5) of Regulation~(EU)~2020/852.

\medskip
\textbf{2.}~The activity meets either of the following criteria:

\begin{enumerate}[label=(\alph*),nosep,leftmargin=2em,itemsep=2pt]
    \item at construction, measurement equipment for monitoring of physical emissions, such as those from methane leakage, is installed or a leak detection and repair programme is introduced;
    \item at operation, physical measurement of emissions are reported and leak is eliminated.
\end{enumerate}

\medskip
\textbf{3.}~Where the activity blends fossil gaseous fuels with gaseous or liquid biofuels, the agricultural biomass used for the production of the biofuels complies with the criteria laid down in Article~29, paragraphs~2 to~5, of Directive~(EU)~2018/2001 while forest biomass complies with the criteria laid down in Article~29, paragraphs~6 and~7, of that Directive.

\medskip
{\scriptsize\textsuperscript{(230)} Regulation (EU) 2018/1999 of the European Parliament and of the Council of 11 December 2018 on the governance of the energy union and climate action, amending regulations (EC) No~663/2009 and (EC) No~715/2009 [...], Council Directives 2009/119/EC and (EU)~2015/652 and repealing Regulation (EU) No~525/2013 (OJ~L~328, 21.12.2018, p.~1).}

\end{tcolorbox}
\caption{Input regulation text for Activity CCM~4.29, rendered as it appears to human readers. Note: the two \texttt{<ol>} elements (a--f and a--g) share no explicit grouping in the HTML source; the semantic pathway split (\texttt{1(a--e)} vs.\ \texttt{1(f)}) must be inferred. The unnumbered paragraphs between criteria~1 and~2 (starting ``Compliance with the criteria\ldots'') have no identifiers in the source.}
\label{fig:input_regulation}
\end{figure}

\clearpage

\begin{tcolorbox}[
  colback=white, colframe=black!50,
  boxrule=0.5pt, arc=2pt,
  left=5pt, right=5pt, top=4pt, bottom=4pt,
  width=\textwidth,
  title={\textbf{Annotation Legend}}
]
\small
\colorbox{pathwayHL}{\strut\;\textbf{Pathway Detection}\;}~%
Inferred structural grouping absent from source HTML.\hfill
\colorbox{footnoteHL}{\strut\;\textbf{Footnote Structuring}\;}~%
Items classified as \texttt{must\_be\_fetched} vs.\ \texttt{citation\_only}.\\[3pt]
\colorbox{threshHL}{\strut\;\textbf{Threshold Inheritance}\;}~%
Values propagated from a parent criterion.\hfill
\colorbox{corrHL}{\strut\;\textbf{Cross-Ref Correction}\;}~%
Erroneous cross-references detected and corrected by the ODR stage.\\[3pt]
\colorbox{anchorHL}{\strut\;\textbf{Anchored Paragraph}\;}~%
Unnumbered source paragraph assigned a synthetic criterion ID by the pipeline.\\[3pt]
\colorbox{enrichHL}{\strut\;\textbf{RAG Enrichment}\;}~%
Referenced content retrieved, summarized, and embedded inline by the RAG sub-pipeline.
\end{tcolorbox}

\medskip

\begin{tcolorbox}[
    title={\textbf{Structured JSON Output --- Page 1 of 4: Root and Low-Emission Pathway 1(a--e)}},
    colback=white,colframe=black!70,fonttitle=\small,
    boxrule=0.4pt,arc=2pt,left=3pt,right=3pt,top=2pt,bottom=2pt,
    width=\textwidth,
    breakable
]
\begin{lstlisting}[language=json,
  basicstyle=\ttfamily\fontsize{6.5pt}{8pt}\selectfont,
  xleftmargin=0pt,
  linewidth=\dimexpr\textwidth-10pt\relax,
  breaklines=true,
  breakatwhitespace=false,
  columns=flexible,
  postbreak=\mbox{\textcolor{gray}{$\hookrightarrow$}\space},
  escapeinside={(*@}{@*)}]
{
  "criterion_id": "ROOT",
  "category": "Qualitative",
  "applicability": "Mandatory",
  "tags": [],
  "rule_summary": "The activity must satisfy all top-level criteria (1, 2, and 3).",
  "evaluation_logic": "AND",
  "n_required": null,
  "threshold": null,
  "references": null,
  "dependencies": null,
  "footnotes": [],
  "sub_criteria": [
    {
      "criterion_id": "1",
      "category": "Qualitative",
      "applicability": "Mandatory",
      "tags": [],
      "rule_summary": "...either the low-emission pathway (a-e) or the transitional pathway (f).",
      (*@\hlP{"evaluation\_logic": "OR",              // Two mutually exclusive pathways}@*)
      "n_required": null,
      "threshold": null,
      "references": null,
      "dependencies": null,
      "footnotes": [],
      "sub_criteria": [
        {
          (*@\hlP{"criterion\_id": "1(a-e)",          // Inferred AND group}@*)
          "category": "Qualitative",
          "applicability": "Mandatory",
          "tags": [],
          (*@\hlP{"rule\_summary": "Groups 1(a)-1(e) as cumulative low-emission conditions forming one alternative pathway.",}@*)
          (*@\hlP{"evaluation\_logic": "AND",}@*)
          "n_required": null,
          "threshold": null,
          "references": null,
          "dependencies": null,
          "footnotes": [],
          "sub_criteria": [
            {
              "criterion_id": "1(a)",
              "category": "Quantitative",    // Quantitative: has threshold
              "applicability": "Mandatory",
              "tags": [],
              "rule_summary": "Life-cycle GHG emissions must be below 100 g CO2e/kWh.",
              "evaluation_logic": "LEAF",
              "n_required": null,
              "threshold": {
                "quantitative": {
                  "logic": null,
                  "items": [
                    { "metric": "Life-cycle GHG emissions",
                      "operator": "<", "value": 100,
                      "unit": "g CO2e/kWh", "period": null }
                  ]
                },
                "temporal": null
              },
              "references": null, "dependencies": null,
              "footnotes": [], "sub_criteria": []
            },
            {
              "criterion_id": "1(b)",
              "category": "Qualitative",
              "applicability": "Mandatory",
              "tags": ["Methodology"],
              "rule_summary": "Emissions calculated using Rec. 2013/179/EU or ISO 14067:2018 or ISO 14064-1:2018.",
              "evaluation_logic": "LEAF",
              "n_required": null,
              "threshold": null,
              "references": {
                "logic": "OR",               // OR-linked external references
                "sources": [
                  { "text": "Recommendation 2013/179/EU",
                    "type": "must_be_fetched", "celex_id": "32013H0179",
                    (*@\hlE{"enrichment": \{}@*)
                      (*@\hlE{"status": "retrieved",}@*)
                      (*@\hlE{"summary": "Provides the PEF method for}@*)
                         (*@\hlE{measuring life-cycle environmental}@*)
                         (*@\hlE{performance including GHG emissions.",}@*)
                      (*@\hlE{"key\_facts": ["PEF life-cycle method",}@*)
                         (*@\hlE{"Environmental performance of products"],}@*)
                      (*@\hlE{"thresholds": [],}@*)
                      (*@\hlE{"confidence": 0.85}@*)
                    (*@\hlE{\}}@*)
                  },
                  { "text": "ISO 14067:2018",
                    "type": "must_be_fetched", "celex_id": null,
                    (*@\hlE{"enrichment": \{"status": "paywalled"\}}@*)
                  },
                  { "text": "ISO 14064-1:2018",
                    "type": "must_be_fetched", "celex_id": null,
                    (*@\hlE{"enrichment": \{"status": "paywalled"\}}@*)
                  }
                ]
              },
              "dependencies": null, "footnotes": [], "sub_criteria": []
            },
            {
              "criterion_id": "1(c)",
              "category": "Qualitative", "applicability": "Mandatory",
              "tags": ["Verification"],
              "rule_summary": "Life-cycle GHG emissions verified by independent third party.",
              "evaluation_logic": "LEAF",
              "n_required": null, "threshold": null, "references": null,
              "dependencies": null, "footnotes": [], "sub_criteria": []
            },
            {
              "criterion_id": "1(d)",
              "category": "Qualitative", "applicability": "Conditional",
              "tags": ["Methodology"],
              "rule_summary": "If abatement is used, it must comply with the relevant Section of this Annex.",
              "evaluation_logic": "LEAF",
              "n_required": null, "threshold": null,
              "references": { "logic": null, "sources": [
                { "text": "the relevant Section of this Annex",
                  "type": "must_be_fetched", "celex_id": null,
                  (*@\hlE{"enrichment": \{"status": "retrieved",}@*)
                    (*@\hlE{"summary": "Covers gas networks for}@*)
                       (*@\hlE{hydrogen and low-carbon gases,}@*)
                       (*@\hlE{including new and converted}@*)
                       (*@\hlE{networks.",}@*)
                    (*@\hlE{"key\_facts": ["Networks for hydrogen}@*)
                       (*@\hlE{or low-carbon gases",}@*)
                       (*@\hlE{"Leak detection for methane"],}@*)
                    (*@\hlE{"thresholds": [],}@*)
                    (*@\hlE{"confidence": 0.85\}}@*)
                }
              ]},
              "dependencies": null, "footnotes": [], "sub_criteria": []
            },
            {
              "criterion_id": "1(e)",
              "category": "Qualitative", "applicability": "Conditional",
              "tags": ["Methodology"],
              "rule_summary": "If CO2 is captured, transport and storage must comply with Sections 5.11 and 5.12.",
              "evaluation_logic": "LEAF",
              "n_required": null, "threshold": null,
              "references": { "logic": "AND", "sources": [
                { "text": "Section 5.11 of this Annex",
                  "type": "must_be_fetched", "celex_id": null,
                  (*@\hlE{"enrichment": \{"status": "retrieved",}@*)
                    (*@\hlE{"summary": "CO2 transport must limit}@*)
                       (*@\hlE{leakage to 0.5\% of mass transported.",}@*)
                    (*@\hlE{"key\_facts": ["CO2 leakage limit",}@*)
                       (*@\hlE{"Monitoring verified by third party"],}@*)
                    (*@\hlE{"thresholds": ["0.5\%"],}@*)
                    (*@\hlE{"confidence": 0.90\}}@*)
                },
                { "text": "Section 5.12 of this Annex",
                  "type": "must_be_fetched", "celex_id": null,
                  (*@\hlE{"enrichment": \{"status": "retrieved",}@*)
                    (*@\hlE{"summary": "Storage sites require}@*)
                       (*@\hlE{characterisation, leakage detection,}@*)
                       (*@\hlE{and compliance with Directive}@*)
                       (*@\hlE{2009/31/EC.",}@*)
                    (*@\hlE{"key\_facts": ["Storage site}@*)
                       (*@\hlE{assessment", "Leakage detection}@*)
                       (*@\hlE{and monitoring"],}@*)
                    (*@\hlE{"thresholds": [],}@*)
                    (*@\hlE{"confidence": 0.90\}}@*)
                }
              ]},
              "dependencies": null, "footnotes": [], "sub_criteria": []
            }
          ]
        },
        // continued on next page: transitional pathway 1(f)
\end{lstlisting}
\end{tcolorbox}
\captionof{figure}{Structured JSON output (Page~1 of~4). The root AND container joins three top-level criteria. Criterion~1 uses OR to select between the two pathways. The \colorbox{pathwayHL}{inferred AND group \texttt{1(a--e)}} does not exist in the source HTML---it is created by semantic pathway detection. Criterion~\texttt{1(b)} demonstrates \colorbox{enrichHL}{inline RAG enrichment}: the EU legal act is \emph{retrieved} with a criterion-conditioned summary, while ISO standards are marked \emph{paywalled}. Criterion~\texttt{1(e)} shows a \emph{retrieved} internal cross-reference with an extracted threshold (\texttt{0.5\%}). Fields \texttt{verbatim\_text}, \texttt{link}, and \texttt{link\_status} are omitted throughout for brevity.}
\label{fig:json_output_1}

\bigskip

\begin{tcolorbox}[
    title={\textbf{Structured JSON Output --- Page 2 of 4: Transitional Pathway 1(f)}},
    colback=white,colframe=black!70,fonttitle=\small,
    boxrule=0.4pt,arc=2pt,left=3pt,right=3pt,top=2pt,bottom=2pt,
    width=\textwidth,
    breakable
]
\begin{lstlisting}[language=json,
  basicstyle=\ttfamily\fontsize{6.5pt}{8pt}\selectfont,
  xleftmargin=0pt,
  linewidth=\dimexpr\textwidth-10pt\relax,
  breaklines=true,
  breakatwhitespace=false,
  columns=flexible,
  postbreak=\mbox{\textcolor{gray}{$\hookrightarrow$}\space},
  escapeinside={(*@}{@*)}]
        // continued: second alternative under criterion "1"
        {
          "criterion_id": "1(f)",            // Transitional pathway
          "category": "Qualitative",
          "applicability": "Mandatory",
          "tags": [],
          "rule_summary": "Facilities permitted by 31 Dec 2030 must comply with all seven sub-criteria.",
          "evaluation_logic": "AND",
          "n_required": null,
          "threshold": {
            "quantitative": null,
            "temporal": {                    // Deadline on parent container
              "logic": null,
              "items": [{ "type": "deadline", "date": "2030-12-31" }]
            }
          },
          "references": null, "dependencies": null, "footnotes": [],
          "sub_criteria": [
            {
              "criterion_id": "1(f)(a)",
              "category": "Quantitative", "applicability": "Mandatory",
              "tags": [],
              "rule_summary": "Direct GHG < 270 g CO2e/kWh, or annual average <= 550 kg CO2e/kW over 20 years.",
              "evaluation_logic": "LEAF",
              "n_required": null,
              "threshold": {
                "quantitative": {
                  "logic": "OR",             // Two alternative limits
                  "items": [
                    { "metric": "Direct GHG emissions",
                      "operator": "<", "value": 270,
                      "unit": "g CO2e/kWh", "period": null },
                    { "metric": "Average annual direct GHG emissions",
                      "operator": "<=", "value": 550,
                      "unit": "kg CO2e/kW",
                      "period": { "type": "bounded", "value": 20, "unit": "years" } }
                  ]
                },
                "temporal": null
              },
              "references": null, "dependencies": null,
              "footnotes": [], "sub_criteria": []
            },
            {
              "criterion_id": "1(f)(b)",
              "category": "Qualitative", "applicability": "Mandatory",
              "tags": ["Assessment"],
              "rule_summary": "Published comparative assessment must show renewables cannot replace the power.",
              "evaluation_logic": "LEAF",
              "n_required": null, "threshold": null, "references": null,
              "dependencies": null, "footnotes": [], "sub_criteria": []
            },
            {
              "criterion_id": "1(f)(c)",
              "category": "Qualitative", "applicability": "Mandatory",
              "tags": ["Replacement"],
              "rule_summary": "Must replace an existing high-emission facility using solid or liquid fossil fuels.",
              "evaluation_logic": "LEAF",
              "n_required": null, "threshold": null, "references": null,
              "dependencies": null, "footnotes": [], "sub_criteria": []
            },
            {
              "criterion_id": "1(f)(d)",
              "category": "Quantitative", "applicability": "Mandatory",
              "tags": ["Replacement"],
              "rule_summary": "New capacity must not exceed replaced capacity by more than 15%.",
              "evaluation_logic": "LEAF",
              "n_required": null,
              "threshold": {
                "quantitative": { "logic": null, "items": [
                  { "metric": "Capacity increase",
                    "operator": "<=", "value": 15, "unit": "%", "period": null }
                ]},
                "temporal": null
              },
              "references": null, "dependencies": null,
              "footnotes": [], "sub_criteria": []
            },
            {
              "criterion_id": "1(f)(e)",
              "category": "Qualitative", "applicability": "Mandatory",
              "tags": ["Commitment"],
              "rule_summary": "Full switch to renewable/low-carbon fuels by 31 Dec 2035 with approved plan.",
              "evaluation_logic": "LEAF",
              "n_required": null,
              "threshold": {
                "quantitative": null,
                "temporal": { "logic": null,
                  "items": [{ "type": "deadline", "date": "2035-12-31" }]
                }
              },
              "references": null, "dependencies": null,
              "footnotes": [], "sub_criteria": []
            },
            {
              "criterion_id": "1(f)(f)",
              "category": "Quantitative", "applicability": "Mandatory",
              "tags": [],
              "rule_summary": "Replacement must reduce lifetime GHG by >= 55%.",
              "evaluation_logic": "LEAF",
              "n_required": null,
              "threshold": {
                "quantitative": { "logic": null, "items": [
                  { "metric": "GHG emission reduction",
                    "operator": ">=", "value": 55, "unit": "%",
                    "period": { "type": "lifetime" } }
                ]},
                "temporal": null
              },
              "references": null, "dependencies": null,
              "footnotes": [], "sub_criteria": []
            },
            {
              "criterion_id": "1(f)(g)",
              "category": "Qualitative", "applicability": "Conditional",
              "tags": ["Commitment"],
              "rule_summary": "If in coal-using Member State, that state must have committed to phase out coal.",
              "evaluation_logic": "LEAF",
              "n_required": null, "threshold": null,
              "references": { "logic": "OR", "sources": [
                { "text": "Article 3 of Regulation (EU) 2018/1999",
                  "type": "must_be_fetched", "celex_id": "32018R1999",
                  (*@\hlE{"enrichment": \{"status": "retrieved",}@*)
                    (*@\hlE{"summary": "Requires each Member State}@*)
                       (*@\hlE{to notify the Commission of an}@*)
                       (*@\hlE{integrated national energy and}@*)
                       (*@\hlE{climate plan.",}@*)
                    (*@\hlE{"key\_facts": ["Integrated national}@*)
                       (*@\hlE{energy and climate plans",}@*)
                       (*@\hlE{"Plans publicly available"],}@*)
                    (*@\hlE{"thresholds": [],}@*)
                    (*@\hlE{"confidence": 0.85\}}@*)
                },
                { "text": "another instrument",
                  "type": "must_be_fetched", "celex_id": null,
                  (*@\hlE{"enrichment": \{"status": "skipped"\}}@*)
                }
              ]},
              "dependencies": null,
              "footnotes": [{
                "id": "fn-230",
                "categories": ["Legal Reference"],
                "items": [
                  { "kind": "EU Legal Act",
                    "title": "Regulation (EU) 2018/1999",
                    "celex_id": "32018R1999",
                    "type": (*@\hlF{"must\_be\_fetched"}@*), "oj": "OJ L 328, 21.12.2018, p. 1",
                    (*@\hlE{"enrichment": \{"status": "retrieved",}@*)
                      (*@\hlE{"summary": "Sets national renewable}@*)
                         (*@\hlE{energy targets within integrated}@*)
                         (*@\hlE{energy and climate plans.",}@*)
                      (*@\hlE{"key\_facts": ["National renewable}@*)
                         (*@\hlE{energy targets", "Commission}@*)
                         (*@\hlE{assesses plan ambition"],}@*)
                      (*@\hlE{"thresholds": [],}@*)
                      (*@\hlE{"confidence": 0.60\}}@*)
                  },
                  { "kind": "EU Legal Act",
                    "title": "Regulation (EC) No 663/2009",
                    "type": (*@\hlF{"citation\_only"}@*) },
                  { "kind": "EU Legal Act",
                    "title": "Directive 94/22/EC",
                    "type": (*@\hlF{"citation\_only"}@*) }
                  // ... +10 more (*@\hlF{"citation\_only"}@*) items omitted
                ],
                "definitions": [], "notes": []
              }],
              "sub_criteria": []
            }
          ]
        },
        // continued on next page: anchored paragraphs
\end{lstlisting}
\end{tcolorbox}
\captionof{figure}{Structured JSON output (Page~2 of~4). Transitional pathway \texttt{1(f)} shows: temporal deadline on the parent container; OR-linked quantitative thresholds with bounded periods on \texttt{1(f)(a)}; a lifetime period on \texttt{1(f)(f)}; \colorbox{footnoteHL}{structured footnote~230 with 13 items distinguishing \texttt{must\_be\_fetched} from \texttt{citation\_only}}, where the \texttt{must\_be\_fetched} item also carries \colorbox{enrichHL}{enrichment}; and \colorbox{enrichHL}{\texttt{1(f)(g)} enrichment} showing a \emph{retrieved} EU legal act alongside a \emph{skipped} vague reference.}
\label{fig:json_output_2}

\bigskip

\begin{tcolorbox}[
    title={\textbf{Structured JSON Output --- Page 3 of 4: Anchored Paragraphs and Corrections}},
    colback=white,colframe=black!70,fonttitle=\small,
    boxrule=0.4pt,arc=2pt,left=3pt,right=3pt,top=2pt,bottom=2pt,
    width=\textwidth,
    breakable
]
\begin{lstlisting}[language=json,
  basicstyle=\ttfamily\fontsize{6.5pt}{8pt}\selectfont,
  xleftmargin=0pt,
  linewidth=\dimexpr\textwidth-10pt\relax,
  breaklines=true,
  breakatwhitespace=false,
  columns=flexible,
  postbreak=\mbox{\textcolor{gray}{$\hookrightarrow$}\space},
  escapeinside={(*@}{@*)}]
        // continued: unnumbered paragraphs anchored under criterion "1"
        {
          (*@\hlA{"criterion\_id": "1(f).Verification",  // Anchored paragraph}@*)
          "category": "Qualitative", "applicability": "Mandatory",
          "tags": ["Verification"],
          "rule_summary": "An independent third party must annually verify compliance with the
             transitional pathway criteria (1(f)) and submit a report to the Commission.
             (*@\hlC{[CORR FROM:1(b) TO:1(f) REASON:crossref]}@*)",
          "evaluation_logic": "AND",
          "n_required": null, "threshold": null, "references": null,
          "dependencies": {
            "condition_summary": "This verification requirement applies only to activities complying via the transitional pathway.",
            "min_conditions_to_meet": 1,
            "clauses": [
              { "criterion_id": "1(f)", "status": "Affirmation" }
            ]
          },
          "footnotes": [],
          "sub_criteria": [
            {
              (*@\hlA{"criterion\_id": "1(f).Verification(a)",}@*)
              "category": "Quantitative", "applicability": "Mandatory",
              "tags": ["Verification"],
              "rule_summary": "Certify direct GHG < 270 g CO2e/kWh or on track for 550 kg CO2e/kW
                 over 20 years. (*@\hlT{[THRESHOLD\_FROM:1(f)(a)]}@*) (*@\hlC{[CORR FROM:1(b)(i) TO:1(f)(a)]}@*)",
              "evaluation_logic": "LEAF",
              "n_required": null,
              "threshold": {
                "quantitative": {
                  "logic": "OR",
                  "items": [
                    { "metric": "Direct GHG emissions",
                      "operator": "<", "value": 270,
                      "unit": "g CO2e/kWh", "period": null },
                    { "metric": "Average annual direct GHG emissions",
                      "operator": "<=", "value": 550,
                      "unit": "kg CO2e/kW",
                      "period": { "type": "bounded", "value": 20, "unit": "years" } }
                  ]
                },
                "temporal": null
              },
              "references": null, "dependencies": null,
              "footnotes": [], "sub_criteria": []
            },
            {
              (*@\hlA{"criterion\_id": "1(f).Verification(b)",}@*)
              "category": "Quantitative", "applicability": "Conditional",
              "tags": ["Verification"],
              "rule_summary": "Assess trajectory to meet 550 kg CO2e/kW threshold.
                 (*@\hlT{[THRESHOLD\_FROM:1(f)(a)]}@*) (*@\hlC{[CORR FROM:1(b)(i) TO:1(f)(a)]}@*)",
              "evaluation_logic": "LEAF",
              "n_required": null,
              "threshold": {
                "quantitative": { "logic": null, "items": [
                  { "metric": "Average annual direct GHG emissions",
                    "operator": "<=", "value": 550,
                    "unit": "kg CO2e/kW",
                    "period": { "type": "bounded", "value": 20, "unit": "years" } }
                ]},
                "temporal": null
              },
              "references": null, "dependencies": null,
              "footnotes": [], "sub_criteria": []
            },
            {
              (*@\hlA{"criterion\_id": "1(f).Verification(c)",}@*)
              "category": "Qualitative", "applicability": "Mandatory",
              "tags": ["Verification"],
              "rule_summary": "Assess credible trajectory to complete fuel switch by 2035.
                 (*@\hlT{[THRESHOLD\_FROM:1(f)(e)]}@*) (*@\hlC{[CORR FROM:1(b)(v) TO:1(f)(e)]}@*)",
              "evaluation_logic": "LEAF",
              "n_required": null,
              "threshold": {
                "quantitative": null,
                "temporal": { "logic": null,
                  "items": [{ "type": "deadline", "date": "2035-12-31" }]
                }
              },
              "references": null, "dependencies": null,
              "footnotes": [], "sub_criteria": []
            }
          ]
        },
        // continued on next page: BackgroundInformation, criteria 2 and 3
\end{lstlisting}
\end{tcolorbox}
\captionof{figure}{Structured JSON output (Page~3 of~4). The \colorbox{anchorHL}{\texttt{1(f).Verification}} anchored paragraph and its sub-criteria are unnumbered in the source---the pipeline assigns them synthetic IDs. \colorbox{corrHL}{\texttt{[CORR]}} corrections fix erroneous ``point~1(b)'' references. \colorbox{threshHL}{\texttt{[THRESHOLD\_FROM]}} propagates values from \texttt{1(f)(a)} and deadlines from \texttt{1(f)(e)}.}
\label{fig:json_output_3}

\bigskip

\begin{tcolorbox}[
    title={\textbf{Structured JSON Output --- Page 4 of 4: Background, Criteria 2--3, and Closing}},
    colback=white,colframe=black!70,fonttitle=\small,
    boxrule=0.4pt,arc=2pt,left=3pt,right=3pt,top=2pt,bottom=2pt,
    width=\textwidth,
    breakable
]
\begin{lstlisting}[language=json,
  basicstyle=\ttfamily\fontsize{6.5pt}{8pt}\selectfont,
  xleftmargin=0pt,
  linewidth=\dimexpr\textwidth-10pt\relax,
  breaklines=true,
  breakatwhitespace=false,
  columns=flexible,
  postbreak=\mbox{\textcolor{gray}{$\hookrightarrow$}\space},
  escapeinside={(*@}{@*)}]
        // continued: second anchored paragraph group under criterion "1"
        {
          (*@\hlA{"criterion\_id": "1(f).BackgroundInformation",}@*)
          "category": "Qualitative", "applicability": "Mandatory",
          "tags": ["BackgroundInformation"],
          "rule_summary": "Groups background context paragraphs for the transitional pathway (1(f)) verification and review.",
          "evaluation_logic": "AND",
          "n_required": null, "threshold": null, "references": null,
          "dependencies": {
            "condition_summary": "Applies to the transitional pathway.",
            "min_conditions_to_meet": 1,
            "clauses": [
              { "criterion_id": "1(f)", "status": "Affirmation" }
            ]
          },
          "footnotes": [],
          "sub_criteria": [
            {
              (*@\hlA{"criterion\_id": "1(f).BackgroundInformation(a)",}@*)
              "category": "Qualitative", "applicability": "Mandatory",
              "tags": ["BackgroundInformation"],
              "rule_summary": "Verifier considers planned vs realised emissions, operating hours,
                 and use of renewable gases. (*@\hlC{[CORR FROM:1(b) TO:1(f)]}@*)",
              "evaluation_logic": "LEAF",
              "n_required": null, "threshold": null, "references": null,
              "dependencies": null, "footnotes": [], "sub_criteria": []
            },
            {
              (*@\hlA{"criterion\_id": "1(f).BackgroundInformation(b)",}@*)
              "category": "Qualitative", "applicability": "Conditional",
              "tags": ["BackgroundInformation"],
              "rule_summary": "Commission may address opinions to operators and considers reports in its Article 19(5) review.",
              "evaluation_logic": "LEAF",
              "n_required": null, "threshold": null,
              "references": { "logic": null, "sources": [
                { "text": "Article 19(5) of Regulation (EU) 2020/852",
                  "type": "must_be_fetched", "celex_id": "32020R0852",
                  (*@\hlE{"enrichment": \{"status": "retrieved",}@*)
                    (*@\hlE{"summary": "Commission must review}@*)
                       (*@\hlE{technical screening criteria at}@*)
                       (*@\hlE{least every three years.",}@*)
                    (*@\hlE{"key\_facts": ["Review every three}@*)
                       (*@\hlE{years", "Assess impact on}@*)
                       (*@\hlE{capital markets"],}@*)
                    (*@\hlE{"thresholds": [],}@*)
                    (*@\hlE{"confidence": 0.85\}}@*)
                }
              ]},
              "dependencies": {
                "condition_summary": "Applies when verification reports are submitted to the Commission.",
                "min_conditions_to_meet": 1,
                "clauses": [
                  { "criterion_id": "1(f).Verification", "status": "Affirmation" }
                ]
              },
              "footnotes": [], "sub_criteria": []
            }
          ]
        }
      ]
    },
    {
      "criterion_id": "2",
      "category": "Qualitative", "applicability": "Mandatory",
      "tags": [],
      "rule_summary": "The activity must implement emission monitoring and repair measures during construction or operation.",
      "evaluation_logic": "OR",
      "n_required": null, "threshold": null, "references": null,
      "dependencies": null, "footnotes": [],
      "sub_criteria": [
        {
          "criterion_id": "2(a)",
          "category": "Qualitative", "applicability": "Mandatory",
          "tags": ["Methodology"],
          "rule_summary": "At construction: install emission monitoring equipment or introduce a leak detection program.",
          "evaluation_logic": "LEAF",
          "n_required": null, "threshold": null, "references": null,
          "dependencies": null, "footnotes": [], "sub_criteria": []
        },
        {
          "criterion_id": "2(b)",
          "category": "Qualitative", "applicability": "Mandatory",
          "tags": ["Methodology"],
          "rule_summary": "At operation: report physical emissions and eliminate any detected leaks.",
          "evaluation_logic": "LEAF",
          "n_required": null, "threshold": null, "references": null,
          "dependencies": null, "footnotes": [], "sub_criteria": []
        }
      ]
    },
    {
      "criterion_id": "3",
      "category": "Qualitative", "applicability": "Conditional",
      "tags": ["Methodology"],
      "rule_summary": "If blending with biofuels: agricultural biomass must meet Art. 29(2-5) and forest biomass Art. 29(6-7) of Directive (EU) 2018/2001.",
      "evaluation_logic": "LEAF",
      "n_required": null, "threshold": null,
      "references": {
        "logic": "AND",
        "sources": [
          { "text": "Article 29, paragraphs 2 to 5, of Directive (EU) 2018/2001",
            "type": "must_be_fetched", "celex_id": "32018L2001",
            (*@\hlE{"enrichment": \{"status": "retrieved",}@*)
              (*@\hlE{"summary": "Defines renewable energy}@*)
                 (*@\hlE{sources including biomass and}@*)
                 (*@\hlE{biogas.",}@*)
              (*@\hlE{"key\_facts": ["Renewable energy}@*)
                 (*@\hlE{sources defined",}@*)
                 (*@\hlE{"Biomass and biogas included"],}@*)
              (*@\hlE{"thresholds": [],}@*)
              (*@\hlE{"confidence": 0.60\}}@*)
          },
          { "text": "Article 29, paragraphs 6 and 7, of that Directive",
            "type": "must_be_fetched", "celex_id": "32018L2001",
            (*@\hlE{"enrichment": \{"status": "retrieved",}@*)
              (*@\hlE{"summary": "Defines renewable energy}@*)
                 (*@\hlE{sources including biomass and}@*)
                 (*@\hlE{biogas.",}@*)
              (*@\hlE{"key\_facts": ["Renewable energy}@*)
                 (*@\hlE{sources defined",}@*)
                 (*@\hlE{"Biomass and biogas included"],}@*)
              (*@\hlE{"thresholds": [],}@*)
              (*@\hlE{"confidence": 0.60\}}@*)
          }
        ]
      },
      "dependencies": null, "footnotes": [], "sub_criteria": []
    }
  ]
}
\end{lstlisting}
\end{tcolorbox}
\captionof{figure}{Structured JSON output (Page~4 of~4). The \colorbox{anchorHL}{\texttt{BackgroundInformation} anchored group} likewise has no numbering in the source. \colorbox{enrichHL}{Enrichment} on \texttt{BackgroundInfo(b)} and Criterion~3 shows \emph{retrieved} summaries with varying confidence levels.}
\label{lst:json_output}
\twocolumn

\section{ODR Self-Correction Details}
  \label{app:odr_trace}

  The \odr{} observation step categorizes detected issues into four types:
  \begin{itemize}[nosep,leftmargin=1.5em]
      \item \emph{Structural}: format violations or missing required fields;
      \item \emph{Semantic}: operator inconsistent with text (e.g., ``$>$'' when text says ``at most'');
      \item \emph{Completeness}: criteria or thresholds present in the source but absent from the output;
      \item \emph{Consistency}: cross-field contradictions (e.g., category marked Quantitative but no threshold
   data).
  \end{itemize}
  Each issue carries a severity level, and confidence is computed via a penalty model that deducts from a base
  score proportionally to issue count and severity, with the final score clamped to $[0, 1]$.
  
\section{Prompt Templates}
\label{app:prompts}
\label{app:prompt_sse}

All prompts instruct the LLM to return JSON-only output.
Formatting boilerplate (``Return ONLY valid JSON'', output schema definitions) is omitted below for brevity; complete prompts are available in our repository.
Prompts are grouped by pipeline stage (\S\ref{subsec:agents}).

\subsection{Stage 2 (Threshold Extractor): Quantitative Threshold Extraction}
\label{app:prompt_quant}

\begin{tcolorbox}[colback=gray!5, colframe=gray!75, title=Stage 2 (Threshold Extractor): Quantitative Threshold Extraction, fonttitle=\bfseries\small, breakable]
\small
\textbf{Task:} Extract numeric performance thresholds (NOT dates/deadlines) from regulatory text.

\textbf{Input:} \texttt{\{text\}}

\medskip
\textbf{Metric Naming:} Use the COMPLETE qualifier chain from source text:
\begin{itemize}[nosep,leftmargin=*]
    \item ``life-cycle GHG emissions'' $\rightarrow$ \texttt{"Life-cycle GHG emissions"} (NOT just ``GHG emissions'')
    \item ``direct GHG emissions'' $\rightarrow$ \texttt{"Direct GHG emissions"}
    \item ``average annual direct GHG emissions'' $\rightarrow$ include ``Average annual''
    \item For CHANGES: ``does not exceed X by more than 15\%'' $\rightarrow$ \texttt{"Capacity increase"}
    \item For REDUCTIONS: ``reduction of at least 55\%'' $\rightarrow$ \texttt{"GHG emission reduction"}
    \item Causal construction: ``X leads to a reduction of Y'' $\rightarrow$ metric is the MEASURED QUANTITY (Y), not the subject (X)
    \item Metric must be a clean noun phrase; always preserve modifiers (life-cycle, direct, annual, average, net, gross)
\end{itemize}

\medskip
\textbf{Operator Detection (Chain-of-Thought):}

\textbf{Step~1 --- Compliance direction:} What values pass? Lower $\rightarrow$ \texttt{<} or \texttt{<=}. Higher $\rightarrow$ \texttt{>} or \texttt{>=}.

\textbf{Step~2 --- Boundary inclusion:}
``lower than'', ``below'' $\rightarrow$ \texttt{<} (exclusive).
``not exceeding'', ``at most'' $\rightarrow$ \texttt{<=} (inclusive).
``at least'', ``minimum'' $\rightarrow$ \texttt{>=} (inclusive).
``exceeds'', ``more than'' $\rightarrow$ \texttt{>} (exclusive).

\textbf{Defaults:} Emissions/costs $\rightarrow$ \texttt{<=}. Contributions/reductions $\rightarrow$ \texttt{>=}.

\medskip
\textbf{Period} (measurement timeframe, NOT deadline):
\texttt{null}: no period.
\texttt{lifetime}: ``over the lifetime''.
\texttt{bounded}: ``over 20 years'' $\rightarrow$ \texttt{\{type: bounded, value: 20, unit: years\}}.
Note: ``life-cycle'' = methodology, NOT lifetime period $\rightarrow$ use \texttt{null}.

\medskip
\textbf{Threshold vs Condition --- Semantic Test:}

Step~1: Does the number constrain WHO/WHAT the rule applies to? $\rightarrow$ YES = CONDITION (skip).\\
Step~2: Does achieving this number = passing the requirement? $\rightarrow$ YES = THRESHOLD (extract).

Examples: ``contribution of 1.5\% of annual revenue'' $\rightarrow$ compliance test $\rightarrow$ THRESHOLD.
``For establishments with over 50 employees'' $\rightarrow$ constrains WHO $\rightarrow$ CONDITION (skip).

\medskip
\textbf{Compound Alternatives:} ``X or Y'' $\rightarrow$ \texttt{logic="OR"}, extract BOTH.

Example: ``emissions below 270g CO2e/kWh, or annual emissions not exceeding 550 kg CO2e/kW over 20 years'' $\rightarrow$ \texttt{logic="OR"}, items=[\{metric: ``Direct GHG emissions'', op: $<$, value: 270, unit: ``g CO2e/kWh''\}, \{metric: ``Average annual direct GHG emissions'', op: $\leq$, value: 550, unit: ``kg CO2e/kW'', period: \{bounded, 20, years\}\}]
\end{tcolorbox}

\subsection{Stage 2 (Threshold Extractor): Temporal Threshold Extraction}
\label{app:prompt_temporal}

\begin{tcolorbox}[colback=gray!5, colframe=gray!75, title=Stage 2 (Threshold Extractor): Temporal Threshold Extraction, fonttitle=\bfseries\small, breakable]
\small
\textbf{Task:} Extract dates, deadlines, and recurring intervals (NOT numeric thresholds) from regulatory text.

\textbf{Input:} \texttt{\{text\}}

\medskip
\textbf{Types:}
\texttt{deadline}: ``by 31 December 2030'' $\rightarrow$ \texttt{\{type: deadline, date: 2030-12-31\}}.
\texttt{effective\_from}: ``from 1 January 2026''.
\texttt{effective\_until}: ``until 31 December 2025''.
\texttt{window}: ``between 2026 and 2030''.
\texttt{recurring}: ``every five years'' $\rightarrow$ \texttt{\{type: recurring, interval\_value: 5, interval\_unit: years\}}.

\medskip
\textbf{Critical Distinctions:}

\resizebox{\linewidth}{!}{%
\begin{tabular}{@{}lll@{}}
\textbf{Pattern} & \textbf{Meaning} & \textbf{Classification} \\
``by 31 December 2030'' & Deadline & temporal $\checkmark$ \\
``every 5 years'' & Recurring interval & temporal $\checkmark$ \\
``over 20 years'' & Measurement period & quantitative.period $\times$ \\
``at construction'' & Phase name & NOT temporal $\times$ \\
``minimum of five years'' & Duration threshold & quantitative $\times$ \\
\end{tabular}}

\medskip
\textbf{Chain-of-Thought:}
Step~1: Scan for years, full dates, interval words.
Step~2: Classify each --- calendar date/deadline $\rightarrow$ extract; measurement period $\rightarrow$ skip; phase name $\rightarrow$ skip.
Step~3: Extract from qualifying clauses too: ``facilities for which permit is granted by 31 December 2030'' $\rightarrow$ deadline.
\end{tcolorbox}


\subsection{Stage 3 (Content Classifier): Evaluation Logic Inference}
\label{app:prompt_logic}

\begin{tcolorbox}[colback=gray!5, colframe=gray!75, title=Stage 3 (Content Classifier): Evaluation Logic Inference, fonttitle=\bfseries\small, breakable]
\small
\textbf{Task:} Determine the logical relationship between items following the chapeau text.

\textbf{Input:} \texttt{\{chapeau\_text\}}

\medskip
\textbf{3-Step Semantic Reasoning} (apply all three steps):

\textbf{Step~1 --- Simultaneity:} Can ALL children be satisfied at the same time?
\begin{itemize}[nosep,leftmargin=*]
    \item If children describe conditions that CAN coexist $\rightarrow$ AND is possible
    \item If children describe MUTUALLY EXCLUSIVE alternatives $\rightarrow$ must be OR
\end{itemize}

\textbf{Step~2 --- Sufficiency:} Is satisfying ONE child sufficient for compliance?
\begin{itemize}[nosep,leftmargin=*]
    \item ONE child = compliant $\rightarrow$ OR
    \item Need more $\rightarrow$ continue to Step~3
\end{itemize}

\textbf{Step~3 --- Completeness:} Are ALL children required?
\begin{itemize}[nosep,leftmargin=*]
    \item ALL required $\rightarrow$ AND (cumulative)
    \item Specific count $\rightarrow$ N\_OF\_K
    \item Default (unclear) $\rightarrow$ AND (regulatory conservative default)
\end{itemize}

\medskip
\textbf{Special Pattern --- Conditional Tiering:}
When the chapeau introduces values followed by a colon, and children represent conditional categories (different values for different situations based on mutually exclusive attributes such as size, count, or type), use OR --- an entity can only be in ONE category at a time.

\medskip
\textbf{Examples:}

\texttt{A:} ``The activity complies with one of the following criteria''\\
Step~1: check $\rightarrow$ Step~2: ``one of'' = ONE is sufficient $\rightarrow$ \textbf{OR}

\texttt{B:} ``The activity meets all of the following requirements''\\
Step~1: can coexist $\rightarrow$ Step~2: ``all of'' = ONE is not sufficient $\rightarrow$ Step~3: ``all'' = ALL required $\rightarrow$ \textbf{AND}

\texttt{C:} ``Facilities for which approval is granted by 31 December 2027''\\
Step~1: scope statement, not alternatives $\rightarrow$ Step~3: requirements apply to these facilities $\rightarrow$ \textbf{AND}

\texttt{D:} ``The activity demonstrates compliance through either of the following pathways''\\
Step~2: ``either of'' = ONE pathway is sufficient $\rightarrow$ \textbf{OR}

\texttt{E:} ``At least equivalent to the applicable Union legislation''\\
Comparison benchmark, not selection between children $\rightarrow$ \textbf{AND}

\texttt{F:} ``The percentage contribution is at least equivalent to:''\\
Children have conditional phrases (``where agreement includes 1 establishment'', ``where includes 2--10'') --- mutually exclusive tiers $\rightarrow$ \textbf{OR}
\end{tcolorbox}

\subsection{Stage 3 (Content Classifier): Applicability Classification}
\label{app:prompt_applicability}

\begin{tcolorbox}[colback=gray!5, colframe=gray!75, title=Stage 3 (Content Classifier): Applicability Classification, fonttitle=\bfseries\small, breakable]
\small
\textbf{Task:} Determine if this criterion is Mandatory or Conditional.

\textbf{Input:} \texttt{\{criterion\_id\}}, \texttt{\{verbatim\_text\}}

\textbf{Default: Mandatory} unless proven otherwise.

\medskip
\textbf{Chain-of-Thought Reasoning:}

\textbf{Step~1 --- Universal Qualifiers} (``where available'', ``where applicable'', ``where feasible''):

Three questions: (1)~Does this phrase apply to ALL instances universally? (2)~Does it modify HOW to comply, not WHETHER to comply? (3)~If removed, does the core obligation still exist?

If ALL THREE YES $\rightarrow$ Mandatory. Example: ``calculated based on project-specific data, \emph{where available}'' --- data source preference, obligation still exists $\rightarrow$ Mandatory.

If any NO $\rightarrow$ Conditional. Example: ``\emph{where applicable}, assessing whether emissions are on trajectory'' --- gates a specific pathway $\rightarrow$ Conditional.

\medskip
\textbf{Step~2 --- ``Where [entity] [verb] X'' pattern:}
\begin{itemize}[nosep,leftmargin=*]
    \item ``Where facilities \emph{incorporate} any form of abatement\ldots'' $\rightarrow$ Conditional
    \item ``Where the CO\textsubscript{2} \emph{is captured} for underground storage\ldots'' $\rightarrow$ Conditional
    \item ``Where the activity \emph{takes place} on territory of a Member State using coal\ldots'' $\rightarrow$ Conditional
\end{itemize}
These describe SITUATIONS that vary between instances $\rightarrow$ ALWAYS Conditional.

\medskip
\textbf{Step~3 --- Semantic Role Analysis:}
\begin{itemize}[nosep,leftmargin=*]
    \item \textbf{Instance filter} $\rightarrow$ Conditional: characteristic that VARIES between instances. Test: Can two compliant instances differ on this condition? YES $\rightarrow$ Conditional.
    \item \textbf{Universal scope} $\rightarrow$ Mandatory: defining characteristic ALL instances share.
\end{itemize}

\textbf{Step~4 --- Parent context:} AND-parent $\rightarrow$ default Mandatory. OR-parent $\rightarrow$ evaluate independently.

\textbf{Step~5 --- Performance vs.\ Applicability:} Performance target (WHAT to achieve, ``emissions below 100g'') $\rightarrow$ Mandatory. Applicability threshold (WHO, ``facilities over 100~MW'') $\rightarrow$ Conditional.

\medskip
\textbf{Temporal context $\neq$ condition:} ``When undertaking X'' = process stage $\rightarrow$ Mandatory. ``When/Where X occurs'' = situation $\rightarrow$ evaluate.

\textbf{Discretionary actions:} ``may'' $\rightarrow$ Conditional. ``shall'' $\rightarrow$ Mandatory.
\end{tcolorbox}

\subsection{Stage 3 (Content Classifier): Tag Assignment and Rule Summary}
\label{app:prompt_tags}

\begin{tcolorbox}[colback=gray!5, colframe=gray!75, title=Stage 3 (Content Classifier): Tag Assignment and Rule Summary, fonttitle=\bfseries\small, breakable]
\small
\textbf{Task:} Assign semantic tags (0--3) and compose a self-contained rule summary.

\textbf{Input:} \texttt{\{criterion\_id\}}, \texttt{\{verbatim\_text\}}, \texttt{\{threshold\_context\}}

\medskip
\textbf{Task A --- Testability Check:}
Step~1: What must the operator DO or ACHIEVE? Step~2: Can an auditor create a PASS/FAIL checklist? If NOT testable $\rightarrow$ tag as BackgroundInformation.

\medskip
\textbf{Task B --- Tags} (3-step semantic test for EACH potential tag):
\begin{enumerate}[nosep,leftmargin=*]
    \item \textsc{audit test}: Can an auditor create a PASS/FAIL checklist?
    \item \textsc{obligation direction}: WHO must act --- the operator, or someone else?
    \item \textsc{compliance impact}: If the operator ignores this, does compliance fail?
\end{enumerate}
ALL THREE must say YES for a requirement tag. If ANY says NO $\rightarrow$ BackgroundInformation.

\smallskip
\textbf{Tag definitions:}
\begin{itemize}[nosep,leftmargin=*]
    \item \textbf{Verification}: Operator must OBTAIN third-party validation. Apply: ``verified by independent third party'', ``certified by accredited body''. Do not apply: pure thresholds without explicit verification mention; ``verifiable plan'' $\rightarrow$ Commitment.
    \item \textbf{Methodology}: Operator must FOLLOW a specific method/standard/equipment. Apply: ``calculated using ISO 14067'', ``measurement equipment is installed''.
    \item \textbf{Commitment}: Operator must have formal plan/declaration approved by management.
    \item \textbf{Assessment}: Operator must conduct internal study/comparison. Apply: ``comparative assessment'', ``stakeholder consultation''.
    \item \textbf{Replacement}: PRIMARY semantic role is replacing existing equipment/facility. Do not apply when ``replacement'' is context for a metric (``leads to 55\% reduction'' $\rightarrow$ metric, not replacement).
    \item \textbf{BackgroundInformation}: NON-BINDING context. Apply when: (1)~no pass/fail test, (2)~subject is NOT the operator, (3)~no compliance impact. Examples: ``verifier takes into account'', ``Commission may address an opinion''.
\end{itemize}

\smallskip
\textbf{Distinction table:}

{\footnotesize
\resizebox{\linewidth}{!}{%
\begin{tabular}{@{}llll@{}}
\textbf{Pattern} & \textbf{Actor} & \textbf{Binding?} & \textbf{Tag} \\
``Compliance IS VERIFIED by\ldots'' & Operator & Yes & Verification \\
``The verifier TAKES INTO ACCOUNT\ldots'' & Verifier & No & BackgroundInfo \\
``IS SUBJECT TO consultation'' & Operator & Yes & Assessment \\
``The Commission MAY address\ldots'' & Commission & No & BackgroundInfo \\
\end{tabular}}}

\textbf{Parent chapeau rule:} Parent introducing sub-criteria (``meets either of:'') $\rightarrow$ tags: [].

\medskip
\textbf{Task C --- Rule Summary:} Self-contained summary (1--2 sentences) a compliance professional can understand without the original text.

Structure by classification: Conditional $\rightarrow$ start with condition. Quantitative $\rightarrow$ embed ALL thresholds with values and units. Methodology $\rightarrow$ name the standard.

Bad: ``The level of direct GHG emissions must be certified.'' (missing thresholds).
Good: ``An independent third party must verify that direct GHG emissions are below 270\,g\,CO\textsubscript{2}e/kWh or on track for the 20-year average of 550\,kg\,CO\textsubscript{2}e/kW.''
\end{tcolorbox}

\subsection{Stage 4 (Reference Extractor): Reference Extraction}
\label{app:prompt_ref}

\begin{tcolorbox}[colback=gray!5, colframe=gray!75, title=Stage 4 (Reference Extractor): Reference Extraction, fonttitle=\bfseries\small, breakable]
\small
\textbf{Task:} Extract compliance references that operators must fetch/follow.

\textbf{Input:} \texttt{\{text\}}

\medskip
\textbf{Rule~1 --- Never abbreviate:} Keep ``paragraphs'' as ``paragraphs'' (not ``p'' or ``paras''). Copy text EXACTLY.

\textbf{Rule~2 --- Internal vs.\ external:}
\begin{itemize}[nosep,leftmargin=*]
    \item ``criterion 1(a)'', ``point (c)'' $\rightarrow$ SKIP (internal cross-reference, goes to dependencies)
    \item ``Article 3'', ``ISO 14067'' $\rightarrow$ EXTRACT (external law/standard)
    \item ``Section of this Annex'', ``Annex A'' $\rightarrow$ EXTRACT (external section/appendix)
\end{itemize}
\textbf{Critical distinction:} ``complies with criteria set out in \emph{Section X}'' $\rightarrow$ REFERENCE (document to fetch). ``Where \emph{criterion 1(a)} applies'' $\rightarrow$ SKIP (dependency).

\textbf{Rule~3 --- Legal basis = ONE reference:} ``plan referred to in Article~3 of Regulation~(EU)~2018/1999'' $\rightarrow$ ONE source (Article is part of where the plan is defined).

\textbf{Rule~4 --- Split when independent:}
\begin{itemize}[nosep,leftmargin=*]
    \item ``Sections 5.11 and 5.12'' $\rightarrow$ TWO independent sections $\rightarrow$ SPLIT
    \item ``Directives 92/43/EEC and 2009/147/EC'' $\rightarrow$ TWO legal acts $\rightarrow$ SPLIT
    \item Multi-subject: ``agricultural biomass complies with [Ref$_1$] while forest biomass complies with [Ref$_2$]'' $\rightarrow$ SPLIT (``while'' = coordination conjunction connecting parallel requirements)
    \item ``Article 3 of Regulation X'' $\rightarrow$ ONE (Article is PART OF the Regulation)
\end{itemize}

\textbf{Rule~5 --- CELEX IDs:} Format: \texttt{3[YEAR][TYPE][NUMBER\_4\_DIGITS]}.
Type codes: R=Regulation, L=Directive, D=Decision, H=Recommendation.
Number is ALWAYS zero-padded to 4 digits: 179 $\rightarrow$ ``0179'', 43 $\rightarrow$ ``0043''.
Examples: ``Recommendation 2013/179/EU'' $\rightarrow$ \texttt{32013H0179}. ``Directive 92/43/EEC'' $\rightarrow$ \texttt{31992L0043}.

\medskip
\textbf{Chain-of-Thought} (7 steps):
(1)~Find candidate references.
(2)~Check if internal $\rightarrow$ skip.
(3)~Determine fetch requirement (\texttt{must\_be\_fetched} vs.\ \texttt{citation\_only}).
(4)~Multi-subject detection (``X complies with A while Y complies with B'').
(5)~Independent reference detection (split ``Sections 5.11 and 5.12'').
(6)~Determine logic (OR: ``or'', ``alternatively''; AND: ``and'', cumulative).
(7)~Copy exact text.

\medskip
\textbf{Example:} ``agricultural biomass complies with Article~29, paragraphs~2 to~5, of Directive~(EU)~2018/2001 while forest biomass complies with Article~29, paragraphs~6 and~7, of that Directive''

$\rightarrow$ \texttt{logic: "AND"}, two sources: [Art.~29(2--5), celex: 32018L2001] and [Art.~29(6--7), celex: 32018L2001]. ``while'' = parallel compliance, ``that Directive'' = anaphoric reference.
\end{tcolorbox}

\subsection{Stage 5 (Dependency Resolver): Cross-Reference Disambiguation}
\label{app:prompt_disambig}

\begin{tcolorbox}[colback=gray!5, colframe=gray!75, title=Stage 5 (Dependency Resolver): Cross-Reference Disambiguation, fonttitle=\bfseries\small, breakable]
\small
\textbf{Task:} Disambiguate an ambiguous cross-reference when multiple criteria could match.

\textbf{Input:} \texttt{\{current\_id\}}, \texttt{\{current\_text\}}, \texttt{\{original\_id\}}, three candidates (original, same-family, parent).

\medskip
\textbf{Step~1 --- Understand context:} What is the current criterion about? (verification, background, requirement?)

\textbf{Step~2 --- Analyze each candidate:}
Candidate~A (\texttt{original\_id}): cross-family reference to a sibling?
Candidate~B (\texttt{same\_family\_id}): same-family sub-criterion?
Candidate~C (\texttt{parent\_id}): about compliance with the parent overall?

\textbf{Step~3 --- Determine intent:}
\begin{itemize}[nosep,leftmargin=*]
    \item If current criterion is a verification/background node, the reference likely means ``verify compliance with the parent criterion's requirements'' $\rightarrow$ choose parent
    \item If current criterion discusses a specific sub-requirement $\rightarrow$ choose same-family
    \item If genuinely cross-referencing another top-level criterion $\rightarrow$ keep original
\end{itemize}

\textbf{Output:} \texttt{\{from: original\_id, to: chosen\_id, reason: "crossref"\}} or \texttt{null}.
\end{tcolorbox}

\subsection{Stage 5 (Dependency Resolver): Cross-Reference Correction}
\label{app:prompt_crossref}

\begin{tcolorbox}[colback=gray!5, colframe=gray!75, title=Stage 5 (Dependency Resolver): Cross-Reference Correction, fonttitle=\bfseries\small, breakable]
\small
\textbf{Task:} Find the correct replacement for a NON-EXISTENT cross-reference.

\textbf{Input:} \texttt{\{current\_id\}}, \texttt{\{current\_text\}}, \texttt{\{missing\_references\}}, \texttt{\{available\_ids\}}, \texttt{\{criteria\_with\_thresholds\}}

\medskip
\textbf{Positional Mapping} (Roman numerals $\rightarrow$ Letters):\\
(i)=a, (ii)=b, (iii)=c, (iv)=d, (v)=e, (vi)=f, (vii)=g.\\
So ``point X(y)(v)'' refers to the 5th child = X(corrected\_parent)(e).

\medskip
\textbf{Semantic Matching Rules:}
When ``point A(b)(v)'' doesn't exist but the correct parent is X(y):
\begin{enumerate}[nosep,leftmargin=*]
    \item Map position: (v) = 5th child = (e)
    \item Check if X(y)(e) exists AND its semantic topic matches the context
    \item If both match $\rightarrow$ use X(y)(e)
\end{enumerate}

Example: Text says ``comply with point 1(b)(v)'' but 1(b)(v) doesn't exist.
Missing 1(b)(v) $\rightarrow$ position (v) = 5th = (e). If correct parent is 1(f), look for 1(f)(e).
Read semantic topic of 1(f)(e): ``fuel switch by 2035'' --- if context is about fuel switching $\rightarrow$ USE 1(f)(e).

\medskip
\textbf{Matching Priority:}
\begin{enumerate}[nosep,leftmargin=*]
    \item \textsc{highest}: Criterion with threshold matching the EXACT TOPIC mentioned in text
    \item Same threshold type (quantitative values, temporal deadlines)
    \item \textsc{lowest}: Same structural family (only if no content match found)
\end{enumerate}

\textbf{Forbidden:} Do NOT use parent fallback; do NOT return a parent/ancestor of the current node; do NOT return same ID as both ``from'' and ``to''.
\end{tcolorbox}

\subsection{Stage 5 (Dependency Resolver): Threshold Inheritance Detection}
\label{app:prompt_inheritance}

\begin{tcolorbox}[colback=gray!5, colframe=gray!75, title=Stage 5 (Dependency Resolver): Threshold Inheritance Detection, fonttitle=\bfseries\small, breakable]
\small
\textbf{Task:} Detect if criterion inherits thresholds from another criterion.

\textbf{Input:} \texttt{\{current\_id\}}, \texttt{\{current\_text\}}, \texttt{\{available\_thresholds\}}

\medskip
\textbf{Chain-of-Thought:}

\textbf{Step~1:} Does text have EXPLICIT threshold delegation?\\
EXPLICIT = [threshold-word] + [delegation-phrase] + [criterion-ID]
\begin{itemize}[nosep,leftmargin=*]
    \item Threshold-words: ``threshold'', ``level'', ``limit'', ``value''
    \item Delegation-phrases: ``referred to in'', ``set out in'', ``specified in''
    \item NOT delegation: ``criteria'', ``requirements'', ``conditions'' (too general). ``comply with X'' without threshold-word (that's a dependency).
\end{itemize}

\textbf{Step~2:} If explicit delegation found, extract source criterion ID.

\textbf{Step~3 --- Mandatory Selector Detection:}
\begin{itemize}[nosep,leftmargin=*]
    \item Period indicators (MUST produce selectors): ``over N years'', ``N-year'', ``N years average'' $\rightarrow$ \texttt{has\_period: true, period\_value: N}
    \item Instant indicators: ``instant'', ``at point of generation'' $\rightarrow$ \texttt{has\_period: false}
    \item No period mentioned $\rightarrow$ \texttt{threshold\_selectors: null} (copy all)
\end{itemize}

\medskip
\textbf{Examples:}

\texttt{1:} ``certifying the level of direct GHG emissions referred to in point X(y)(a)''\\
Step~1: ``level'' + ``referred to in'' + ``X(y)(a)'' = EXPLICIT. Step~3: no period $\rightarrow$ null.\\
$\rightarrow$ \texttt{\{threshold\_from: "X(y)(a)", threshold\_selectors: null\}}

\texttt{2:} ``comply with the average threshold over 20 years referred to in point A(b)(c)''\\
Step~1: EXPLICIT. Step~3: ``over 20 years'' $\rightarrow$ selector.\\
$\rightarrow$ \texttt{\{threshold\_from: "A(b)(c)", threshold\_selectors: [\{has\_period: true, period\_value: 20\}]\}}

\texttt{3:} ``Quantified life-cycle GHG emissions are verified by an independent third party''\\
Step~1: No threshold-word + delegation-phrase + criterion-ID pattern $\rightarrow$ \texttt{\{\}}

\texttt{4:} ``Compliance with the criteria set out in this Section is verified''\\
Step~1: ``criteria'' is NOT a threshold-word (too general) $\rightarrow$ \texttt{\{\}}
\end{tcolorbox}

\subsection{Stage 5 (Dependency Resolver): Dependency Detection}
\label{app:prompt_dependency}

\begin{tcolorbox}[colback=gray!5, colframe=gray!75, title=Stage 5 (Dependency Resolver): Dependency Detection, fonttitle=\bfseries\small, breakable]
\small
\textbf{Task:} Detect applicability dependencies --- when one criterion applies ONLY IF another is met/chosen.

\textbf{Input:} \texttt{\{current\_id\}}, \texttt{\{current\_text\}}, \texttt{\{available\_ids\}}

\medskip
\textbf{Semantic Principle --- Three-way Distinction:}

\textbf{DEPENDENCIES} (what we detect): Applicability gating --- ``This criterion applies ONLY IF criterion X is met/chosen.''

\textbf{NOT dependencies --- REFERENCES} (Agent~4 handles these): ``complies with criteria set out in \emph{Section X}'' $\rightarrow$ points to document to fetch, not conditional applicability.

\textbf{NOT dependencies --- Structural/informational}: ``as defined in criterion X'' $\rightarrow$ informational, not conditional.

\medskip
\textbf{Special Case --- Pathway-Specific Anchors:}
Semantic anchor nodes (\texttt{.Verification}, \texttt{.BackgroundInformation}) anchored to a SPECIFIC PATHWAY within an OR-alternative group SHOULD have a dependency on that pathway.

Example: \texttt{1(f).Verification} is anchored to pathway \texttt{1(f)} (one of several OR alternatives under \texttt{1}). Verification applies ONLY when activity uses pathway 1(f) $\rightarrow$ dependency on \texttt{1(f)}.

Counter-example: \texttt{5.Verification} anchored to top-level \texttt{5} (not an OR alternative) $\rightarrow$ applies universally $\rightarrow$ NO dependency.

\medskip
\textbf{KEY TEST:} ``Does this criterion apply ONLY to activities that meet another criterion?'' YES $\rightarrow$ Dependency. NO $\rightarrow$ NOT a dependency.

\medskip
\textbf{Detection Steps:}
(1)~Identify conditional language (``where'', ``if'', ``when'').
(2)~Extract the condition.
(3)~Map condition to criterion ID from available list.
(4)~Validate it's applicability gating (not structural/informational).
(5)~Return dependency or null.

\medskip
\textbf{Examples:}

\texttt{1:} ``For facilities granted permit by 31 December 2030, capacity does not exceed 15\%''\\
``31 December 2030'' is a DATE, not a criterion ID. Inline temporal condition $\rightarrow$ \textbf{null}.

\texttt{2:} ID: \texttt{5.Verification}. ``At beginning of activity, compliance is controlled by authorities''\\
\texttt{5} is top-level section (NOT OR-alternative child) $\rightarrow$ applies universally $\rightarrow$ \textbf{null}.

\texttt{3:} ID: \texttt{1(f).Verification}. ``Where the activity uses pathway 1(f), emissions must be verified''\\
\texttt{1(f)} is child of \texttt{1} in OR group $\rightarrow$ \textbf{dependency} on \texttt{1(f)}.

\texttt{4:} ID: \texttt{1(f).BackgroundInformation(a)}. ``The Commission shall take those reports into account''\\
Anchored to pathway \texttt{1(f)} (OR child) $\rightarrow$ \textbf{dependency} on \texttt{1(f)}.
\end{tcolorbox}

\subsection{Stage 6 (Footnote Processor): Footnote Processing}
\label{app:prompt_footnote}

\begin{tcolorbox}[colback=gray!5, colframe=gray!75, title=Stage 6 (Footnote Processor): Footnote Processing, fonttitle=\bfseries\small, breakable]
\small
\textbf{Task:} Process regulatory footnote and extract ALL referenced documents.

\textbf{Input:} \texttt{\{footnote\_text\}}

\medskip
\textbf{Document Types} (\texttt{kind}):
EU Legal Act $\mid$ International Convention/Treaty $\mid$ Standard $\mid$ Official Guideline/Manual $\mid$ Member-State Plan/Programme $\mid$ Other.

\medskip
\textbf{CELEX ID Construction} (EU acts only):
Format: \texttt{3[YEAR][TYPE][NUMBER]} (10 characters).
Type codes: R=Regulation, L=Directive, D=Decision, H=Recommendation.

Year extraction: ``YYYY/NNNN'' (modern) $\rightarrow$ first part = year. ``NN/YY'' (old) $\rightarrow$ second part = year. IGNORE adoption dates (``of 9 December 1996'').

Examples: ``Regulation (EU) 2018/1999'' $\rightarrow$ \texttt{32018R1999}. ``Directive 98/70/EC'' $\rightarrow$ \texttt{31998L0070}. ``Council Regulation (EC) 338/97'' $\rightarrow$ \texttt{31997R0338}.

\medskip
\textbf{Type Determination:}
\texttt{must\_be\_fetched}: needed for compliance (``calculated according to'', ``complies with'').
\texttt{citation\_only}: background only (``amending'', ``repealing'', ``establishing framework'').

\textbf{Splitting Rule:} Each legal act is SEPARATE: ``amending Regulations 663/2009 and 715/2009, Directives 94/22/EC, 98/70/EC'' $\rightarrow$ 4 items.

\medskip
\textbf{Categories} (multi-label):
\textbf{Legal Reference}: ONLY when footnote cites binding legislation. NOT for ISO/EN standards.
\textbf{Technical Definition}: defines terms.
\textbf{Official Guideline/Standard}: ISO, IPCC, EN standards.
\textbf{Explanatory Note}: clarifications, scope context.

Validation: If NO items have kind=``EU Legal Act'' or ``Treaty'' $\rightarrow$ ``Legal Reference'' MUST NOT be in categories.
\end{tcolorbox}

\subsection{Stage 7 (Schema Assembler): Rule Summary Generation}
\label{app:prompt_summary}

\begin{tcolorbox}[colback=gray!5, colframe=gray!75, title=Stage 7 (Schema Assembler): Rule Summary Generation, fonttitle=\bfseries\small, breakable]
\small
\textbf{Task:} Generate a self-contained rule summary (1--2 sentences) using semantic understanding.

\textbf{Input:} \texttt{\{verbatim\_text\}}, \texttt{\{category\}}, \texttt{\{applicability\}}, \texttt{\{tags\}}, \texttt{\{threshold\}}, \texttt{\{references\}}

\medskip
\textbf{Structure by Classification:}
\begin{itemize}[nosep,leftmargin=*]
    \item \textbf{Conditional:} Start with condition FIRST (``If\ldots'', ``Where\ldots''), then state requirement
    \item \textbf{Core Requirement:} WHO must do WHAT --- specific and actionable
    \item \textbf{Quantitative:} Embed ALL numeric thresholds with values + units. DO: ``must be below 150\,g\,CO\textsubscript{2}e/kWh''. DON'T: ``must meet the threshold''. Multiple (OR): include ALL alternatives. Period-based: include time period.
    \item \textbf{Temporal:} Embed dates directly (``by 31 December 2027'')
    \item \textbf{Methodology/Verification:} Name specific standard or verifier role
\end{itemize}

\textbf{Bracket Notations:}
Threshold inheritance: append \texttt{[THRESHOLD\_FROM:ID]}.
Cross-reference corrections: append \texttt{[CORR FROM:X TO:Y REASON:crossref]}.

\medskip
\textbf{Examples:}

Quantitative: ``Life-cycle GHG emissions must be below 80\,g\,CO\textsubscript{2}e/kWh.''

Conditional + Methodology: ``If the activity blends primary fuels with alternative fuels, the alternative source must comply with sustainability criteria from Directive~(EU)~2018/2001: agricultural biomass must meet Article~29(2--5) and forest biomass must meet Article~29(6--7).''

Quantitative + Temporal + Verification: ``Direct GHG emissions must be below 150\,g\,CO\textsubscript{2}e/kWh or the 15-year average must not exceed 400\,kg\,CO\textsubscript{2}e/kW, with compliance verified by 31~December~2027.''
\end{tcolorbox}

\subsection{ODR: Observe Step}
\label{app:prompt_observe}

\begin{tcolorbox}[colback=gray!5, colframe=gray!75, title=ODR: Observe Step, fonttitle=\bfseries\small, breakable]
\small
\textbf{Task:} Compare the extracted output against the source HTML and enumerate discrepancies.

\textbf{Input:} \texttt{\{html\}}, \texttt{\{output\}}

\textbf{Issue taxonomy:}
\begin{itemize}[nosep,leftmargin=*]
    \item \textbf{Structural}: Format violations, missing required fields, malformed IDs
    \item \textbf{Semantic}: Operator inconsistent with text (e.g., ``$>$'' when text says ``at most'')
    \item \textbf{Completeness}: Criteria or thresholds present in source but missing from output
    \item \textbf{Consistency}: Cross-field contradictions (e.g., parent--child category mismatch)
\end{itemize}

For each issue, report: \texttt{\{type, severity, field, description, source\_evidence\}}
\end{tcolorbox}

\subsection{ODR: Diagnose Step}
\label{app:prompt_diagnose}

\begin{tcolorbox}[colback=gray!5, colframe=gray!75, title=ODR: Diagnose Step, fonttitle=\bfseries\small, breakable]
\small
\textbf{Task:} Analyze the observed issues and determine root causes.

\textbf{Input:} \texttt{\{issues\}}, \texttt{\{output\}}, \texttt{\{history\}} (previous attempts)

\textbf{Produce a DiagnosisResult with:}
\begin{itemize}[nosep,leftmargin=*]
    \item \textbf{root\_cause}: Primary reason for the errors
    \item \textbf{contributing\_factors}: Secondary causes
    \item \textbf{recommended\_action}: One of \{\textsc{retry\_modified}, \textsc{decompose}, \textsc{fallback}, \textsc{accept}, \textsc{escalate}\}
    \item \textbf{specific\_guidance}: Fields to focus on, mistakes to avoid, modifications to apply
\end{itemize}
\end{tcolorbox}

\subsection{RAG: Query Rewriting with Context}
\label{app:prompt_rewrite}

\begin{tcolorbox}[colback=gray!5, colframe=gray!75, title=RAG: Criterion-Conditioned Query Rewriting, fonttitle=\bfseries\small, breakable]
\small
\textbf{System:} You are an EU regulatory retrieval expert. Reformulate EU Taxonomy criterion text into a natural-language question optimized for finding relevant passages in EU regulatory documents.

\textbf{Task:} Reformulate the criterion below into a retrieval-optimized question incorporating activity context.

\textbf{Input:} \texttt{\{query\}}, \texttt{\{activity\_name\}}, \texttt{\{objective\}}, \texttt{\{criteria\_section\}}, \texttt{\{article\_ref\}}

\medskip
\textbf{Instructions:}
\begin{enumerate}[nosep,leftmargin=*]
    \item EXPAND all abbreviations/acronyms to full EU regulatory forms
    \item FRAME as: ``For the [substantial contribution / DNSH] criteria of [activity] under [objective], what does [regulation] require regarding [topic]?''
    \item INCLUDE the article/section reference naturally if provided
    \item INCLUDE key regulatory concepts: requirements, thresholds, definitions, conditions
    \item KEEP numeric values, units, and article references exact
    \item NEVER invent references: only mention Articles/Appendices/Annexes that appear in the criterion text
\end{enumerate}

\medskip
\textbf{Examples:}

Activity: Electricity generation using solar PV. Objective: mitigation. Section: substantial contribution.\\
Criterion: ``The activity complies with the emission threshold set in Article~29(4)(a)''\\
$\rightarrow$ ``For the substantial contribution of electricity generation using solar photovoltaic technology to climate change mitigation, what emission threshold does Article~29, paragraph~4, points (a) establish, including the specific CO\textsubscript{2} limit values and measurement conditions?''

Activity: Afforestation. Objective: adaptation. Section: DNSH.\\
Criterion: ``prevention and avoidance of introduction of invasive alien species''\\
$\rightarrow$ ``For the DNSH criteria of afforestation under climate change adaptation, what measures does EU regulation require to prevent the introduction and spread of invasive alien species, including species of Union concern, containment conditions, and risk assessment?''
\end{tcolorbox}

\subsection{RAG: Retrieval Evaluation}
\label{app:prompt_eval}

\begin{tcolorbox}[colback=gray!5, colframe=gray!75, title=RAG: Retrieval Relevance Evaluation, fonttitle=\bfseries\small, breakable]
\small
\textbf{System:} You are an EU regulatory document analysis expert. Evaluate whether passages contain the substantive requirements relevant to the criterion. Focus on CONTENT --- the actual rules, thresholds, definitions, and conditions --- not on whether a specific article number appears verbatim.

\textbf{Task:} Evaluate whether retrieved passages contain the substantive content needed.

\textbf{Input:} \texttt{\{criterion\}}, \texttt{\{celex\_id\}}, \texttt{\{passages\}}

\medskip
\textbf{Scoring rubric:}
\begin{itemize}[nosep,leftmargin=*]
    \item 0.9--1.0: Actual regulatory requirements, thresholds, definitions, or conditions
    \item 0.7--0.89: Related requirements but missing specific details
    \item 0.5--0.69: Correct regulation but different topic
    \item Below 0.5: Unrelated
\end{itemize}

Do NOT penalize passages for lacking a specific article/section heading. A passage containing the actual emission threshold or definition is highly relevant even without an ``Article~X'' heading.

\textbf{Output:} \texttt{\{confidence, relevant\_passages, gaps, reasoning\}}
\end{tcolorbox}

\subsection{RAG: Query Refinement}
\label{app:prompt_refine}

\begin{tcolorbox}[colback=gray!5, colframe=gray!75, title=RAG: Iterative Query Refinement, fonttitle=\bfseries\small]
\small
\textbf{System:} You are a search query generator. Output ONLY a single search query sentence. No reasoning, no explanation, no first person, no preamble.

\textbf{Task:} Given the original query and retrieval gaps, generate a better, more specific search query.

\textbf{Input:} \texttt{\{original\_query\}}, \texttt{\{gaps\}}

Focus on the specific legal terms, article numbers, or thresholds mentioned in the gaps.
\end{tcolorbox}

\subsection{RAG: Safe Summarization}
\label{app:prompt_summarize}

\begin{tcolorbox}[colback=gray!5, colframe=gray!75, title=RAG: Safe Summarization, fonttitle=\bfseries\small, breakable]
\small
\textbf{System:} You are an EU regulatory compliance expert. Summarize relevant passages from EU legal documents in the context of a specific Technical Screening Criterion.

\textbf{Rules:}
\begin{enumerate}[nosep,leftmargin=*]
    \item Focus on requirements, thresholds, dates, and conditions relevant to the criterion
    \item Quote key thresholds and dates VERBATIM (in quotation marks)
    \item Keep each direct quote under 100 words
    \item Identify specific articles/sections that apply
    \item Note any conditions or exceptions
    \item Be factual --- do NOT invent requirements not in the source text
\end{enumerate}

\textbf{Input:} \texttt{\{criterion\}}, \texttt{\{celex\_id\}}, \texttt{\{passages\}}

\textbf{Output:} \texttt{\{text, key\_facts, thresholds, confidence\}}
\end{tcolorbox}

\subsection{Semantic Equivalence Judge Prompts}
\label{app:judge_prompts}

The following prompt templates are used by the GPT-4o LLM judge (max 150 tokens) to score semantic equivalence between system-extracted and gold-annotated fields.
Each prompt receives three inputs: the \emph{verbatim regulatory text} (ground truth source), the \emph{gold} (human-annotated) field, and the \emph{system} (pipeline-extracted) field.
Scores range from 0 to 5, where \textbf{0}~indicates concordant absence (both sides are null/empty, automatically assigned without an LLM call) and \textbf{1--5} follow the rubric below.
Reported statistics in Table~\ref{tab:main_results} are computed over scores $> 0$ only, i.e., on the 1--5 scale.
All four prompts share the same structure: schema definition, numbered evaluation rules with point deductions, scoring rubric with concrete examples, and JSON output format.
We present the threshold prompt in full and abbreviate the remaining three to their schema and rubric sections; evaluation rules follow the same pattern of field-level matching with graduated penalties.

\begin{tcolorbox}[colback=gray!5, colframe=gray!75, title=Threshold Equivalence Judge (full prompt), fonttitle=\bfseries\small, breakable]
\small
You are an expert evaluator for EU Taxonomy regulatory criteria extraction. Compare the SYSTEM-extracted threshold against the GOLD (human-annotated) threshold for a single criterion node.

\medskip
\textbf{Schema.} Each threshold object has two sub-fields:
\begin{itemize}[nosep,leftmargin=*]
    \item \texttt{quantitative}: \{logic: AND|OR|null, items: [\{metric, operator, value, unit, period\}]\} or null
    \item \texttt{temporal}: \{logic: AND|OR|null, items: [\{type: deadline|effective\_from|window|recurring, \ldots\}]\} or null
\end{itemize}

\medskip
\textbf{Evaluation Rules:}
\begin{enumerate}[nosep,leftmargin=1.5em]
    \item \emph{Structural placement is irrelevant}: recurring intervals under ``temporal'' vs.\ ``quantitative'' are not penalized if semantic content is identical.
    \item \emph{Metric label is free-text}: ``Primary Energy Demand'' vs.\ ``PED'' are equivalent; judge whether they refer to the same physical quantity.
    \item \emph{Operator equivalence}: $\geq$ and ``at least'' are the same; $>$ vs.\ $\geq$ \emph{is} a meaningful difference ($-1$ pt).
    \item \emph{Value must be numerically identical}: 10 vs.\ 10.0 is fine; 10 vs.\ 100 is wrong.
    \item \emph{Unit normalization}: ``\%'' vs.\ ``percent'', ``g CO2e/kWh'' vs.\ ``gCO2eq/kWh'' are equivalent; ``g'' vs.\ ``kg'' is wrong.
    \item \emph{Logic field}: AND vs.\ OR with $\geq 2$ items is a critical error ($-2$ pts). With 1 item, logic=null is expected.
    \item \emph{Period field}: missing but inferable from context is minor ($-1$ pt); wrong value or unit is major ($-2$ pts).
    \item \emph{One-sided presence}: if one side is null, check verbatim text for an explicit numeric constraint.
\end{enumerate}

\medskip
\textbf{Inputs:} \texttt{\{verbatim\}}, \texttt{\{gold\}}, \texttt{\{system\}}

\medskip
\textbf{Scoring Rubric (0--5):}
\begin{itemize}[nosep,leftmargin=*]
    \item \textbf{5 --- Full equivalence}: All items match on value, operator, unit, metric concept, and temporal constraints. Surface differences in label wording or sub-field placement are acceptable.
    \item \textbf{4 --- One minor field-level error}: All items present with correct value/operator/unit, but exactly one minor issue (vague metric label, missing period when inferable, null logic with $\geq 2$ correct items).
    \item \textbf{3 --- Partial capture, one substantive gap}: Primary constraint correct, but one item entirely missing from a multi-item set, or operator direction wrong on one item, or temporal condition absent.
    \item \textbf{2 --- Significant errors}: Threshold recognized but critical details wrong (wrong numeric value, fundamentally different unit, only 1 of $3+$ items, AND/OR inverted).
    \item \textbf{1 --- Wrong or unjustified}: False positive (extraction where text is qualitative) or false negative (null where text has explicit limits).
    \item \textbf{0 --- Not applicable}: Both sides null/empty; concordant absence.
\end{itemize}

\medskip
\textbf{Output:} \texttt{\{"score": <int 0-5>, "reason": "<one sentence>"\}}
\end{tcolorbox}

\begin{tcolorbox}[colback=gray!5, colframe=gray!75, title=Reference Equivalence Judge (abbreviated), fonttitle=\bfseries\small, breakable]
\small
\textbf{Schema.} Each references object has: \texttt{logic} (AND|OR|null), \texttt{sources}: [\{text, type: must\_be\_fetched|citation\_only, link, link\_status, celex\_id\}].

\medskip
\textbf{Key Evaluation Rules:} Sources are matched by CELEX ID, directive/regulation number, or standard identifier (not by text wording). Type mismatches (must\_be\_fetched vs.\ citation\_only) incur $-1$ pt each. Missing sources incur $-1$ pt each. Extra sources present in the verbatim text are acceptable; those absent from the text are false positives ($-1$ pt). Logic field AND vs.\ OR with $\geq 2$ sources: $-1$ pt if wrong.

\medskip
\textbf{Scoring Rubric (0--5):}
\begin{itemize}[nosep,leftmargin=*]
    \item \textbf{5}: Same set of legal acts/standards, correct type classification, correct logic.
    \item \textbf{4}: All sources identified; exactly one minor discrepancy (one type mismatch, or null logic with $\geq 2$ sources).
    \item \textbf{3}: One source missing or one false positive.
    \item \textbf{2}: Two or more sources missing, or $>$50\% type mismatches.
    \item \textbf{1}: Most gold sources absent, or entirely wrong legal acts cited.
    \item \textbf{0}: Both sides null/empty; concordant absence.
\end{itemize}
\end{tcolorbox}

\begin{tcolorbox}[colback=gray!5, colframe=gray!75, title=Footnote Equivalence Judge (abbreviated), fonttitle=\bfseries\small, breakable]
\small
\textbf{Schema.} Each footnote has: \texttt{id}, \texttt{verbatim} (primary content from HTML popover), \texttt{categories} (Legal Reference | Technical Definition | Official Guideline/Standard | Explanatory Note), \texttt{items} (structured references with kind, celex\_id, type), \texttt{definitions} [\{term, definition\}], \texttt{notes}.

\medskip
\textbf{Key Evaluation Rules:} Footnotes are matched by \emph{content} (verbatim field), not by ID. Minor whitespace/encoding differences are acceptable; truncation ($>$20\% missing) is substantive. Category mismatches: extra secondary label $-0.5$ pt, wrong primary category $-1$ pt. Missing parsed items when footnote cites legal acts: $-1$ pt. Missing definitions when footnote defines a term: $-1$ pt.

\medskip
\textbf{Scoring Rubric (0--5):}
\begin{itemize}[nosep,leftmargin=*]
    \item \textbf{5}: Same footnote set (by content); categories, items, definitions all agree.
    \item \textbf{4}: All footnotes present; exactly one minor field error (extra category label, wrong item kind, slightly reworded definition).
    \item \textbf{3}: One footnote missing, or one with truncated content, or items/definitions array empty when clearly needed.
    \item \textbf{2}: Two or more footnotes missing, or fundamentally different verbatim content.
    \item \textbf{1}: Most footnotes missing or wrong content; would mislead a user.
    \item \textbf{0}: Both sides have empty lists; concordant absence.
\end{itemize}
\end{tcolorbox}

\begin{tcolorbox}[colback=gray!5, colframe=gray!75, title=Dependency Equivalence Judge (abbreviated), fonttitle=\bfseries\small, breakable]
\small
\textbf{Schema.} Each dependencies object has: \texttt{condition\_summary} (plain-language description), \texttt{min\_conditions\_to\_meet} (int; 1=OR, len(clauses)=AND), \texttt{clauses}: [\{criterion\_id, status: Affirmation|Negation\}]. ``Affirmation'' = target must be met; ``Negation'' = target must NOT be met.

\medskip
\textbf{Key Evaluation Rules:} Clauses matched by criterion\_id (minor format variations acceptable: ``2.1'' = ``2.1.''). Status mismatch (Affirmation vs.\ Negation) is a major error ($-2$ pts) as it inverts dependency logic. min\_conditions\_to\_meet wrong with $\geq 2$ clauses: $-2$ pts (changes AND/OR semantics). condition\_summary judged by logical meaning, not wording; wrong criterion reference: $-1$ pt.

\medskip
\textbf{Scoring Rubric (0--5):}
\begin{itemize}[nosep,leftmargin=*]
    \item \textbf{5}: All clauses match on criterion\_id, status, and min\_conditions\_to\_meet; condition\_summary conveys same meaning.
    \item \textbf{4}: Correct structure; exactly one minor discrepancy (vague summary, minor ID format difference, min\_conditions differs with only 1 clause).
    \item \textbf{3}: Primary dependency correct; one secondary clause missing, or min\_conditions wrong with $\geq 2$ clauses.
    \item \textbf{2}: Dependency recognized but key error: status inverted, or criterion\_id points to wrong node.
    \item \textbf{1}: False negative (null when text has conditional language) or false positive (fabricated dependencies).
    \item \textbf{0}: Both sides null; concordant absence.
\end{itemize}
\end{tcolorbox}

\subsection{RAG Quality Judge Prompts}
\label{app:rag_prompts}

The following prompt templates are used by the GPT-4o judge (max 200 tokens) to score RAG summary quality.
Each prompt receives four inputs: the \emph{regulatory criterion} being enriched, the \emph{retrieved source passages} from the referenced CELEX document, the \emph{generated summary}, and the \emph{extracted key facts and thresholds}.
Scores range from 1 to 5 following a graduated rubric; the judge returns a JSON object with a score and one-sentence justification.

\begin{tcolorbox}[colback=gray!5, colframe=gray!75, title=Faithfulness Judge, fonttitle=\bfseries\small, breakable]
\small
You are an expert evaluator for regulatory RAG pipeline outputs.
Assess whether the SUMMARY faithfully represents the SOURCE PASSAGES.

\medskip
\textbf{Inputs:} \texttt{\{criterion\_text\}}, \texttt{\{chunk\_texts\}}, \texttt{\{summary\_text\}}, \texttt{\{key\_facts\}}, \texttt{\{thresholds\}}.

\medskip
\textbf{Evaluation Rules:}
\begin{enumerate}[nosep,leftmargin=1.5em]
    \item Faithfulness = grounding. Every claim in the summary must be traceable to the source passages. A claim that is true in general EU law but not in these specific passages is hallucination.
    \item Numeric thresholds in the summary or thresholds list must appear (exactly or equivalently) in the source passages.
    \item Each key fact must be supported by at least one passage.
    \item Omission of passage content is acceptable---only fabrication is penalized.
    \item Article/section references cited in the summary must appear in the passages.
\end{enumerate}

\medskip
\textbf{Scoring Rubric (1--5):}
\begin{itemize}[nosep,leftmargin=*]
    \item \textbf{5 --- Fully faithful}: Every claim, threshold, and key fact is directly supported by the passages. Paraphrasing is acceptable if meaning is preserved.
    \item \textbf{4 --- One minor unsupported detail}: All major claims are grounded. Exactly one minor detail lacks direct support: a slightly overstated paraphrase, a vague article reference, or a key fact that combines two passage ideas in a way not explicitly stated.
    \item \textbf{3 --- One substantive unsupported claim}: Core summary is grounded, but one substantive claim is not supported: a threshold value not in the passages, a key fact that conflates distinct requirements, or a specific regulatory reference not found in the retrieved text.
    \item \textbf{2 --- Multiple unsupported claims}: Two or more claims lack passage support, or a critical threshold is hallucinated. The summary would mislead a compliance assessor on material points.
    \item \textbf{1 --- Mostly hallucinated}: The majority of summary content cannot be traced to the passages. The summary appears drawn from general knowledge rather than retrieved text.
\end{itemize}

\medskip
\textbf{Output:} \texttt{\{"score": <int 1-5>, "reason": "<one sentence>"\}}
\end{tcolorbox}

\begin{tcolorbox}[colback=gray!5, colframe=gray!75, title=Relevance Judge, fonttitle=\bfseries\small, breakable]
\small
You are an expert evaluator for regulatory RAG pipeline outputs.
Assess whether the SUMMARY is relevant to the regulatory CRITERION that triggered the retrieval.

\medskip
\textbf{Inputs:} \texttt{\{criterion\_text\}}, \texttt{\{chunk\_texts\}}, \texttt{\{summary\_text\}}, \texttt{\{key\_facts\}}, \texttt{\{thresholds\}}.

\medskip
\textbf{Evaluation Rules:}
\begin{enumerate}[nosep,leftmargin=1.5em]
    \item Classify each summary claim into three tiers: \emph{directly relevant} (addresses a specific requirement of the criterion), \emph{contextual} (provides useful background but does not directly address the criterion), or \emph{unrelated} (no connection to the criterion).
    \item The criterion text defines relevance---not the passages. A claim may be faithful to the passages but irrelevant to the criterion.
    \item General legal context is ``contextual'' unless it directly clarifies a requirement in the criterion.
    \item Key facts should each relate to a specific aspect of the criterion.
\end{enumerate}

\medskip
\textbf{Scoring Rubric (1--5):}
\begin{itemize}[nosep,leftmargin=*]
    \item \textbf{5 --- Fully relevant}: Every claim in the summary directly addresses a requirement or condition stated in the criterion. Key facts map to specific criterion aspects.
    \item \textbf{4 --- One contextual claim}: All major claims are directly relevant. Exactly one claim provides background context rather than addressing the criterion directly.
    \item \textbf{3 --- Multiple contextual or one unrelated}: Several claims provide only background context, or one claim is entirely unrelated to the criterion.
    \item \textbf{2 --- Significant off-topic content}: The summary contains substantial content unrelated to the criterion, diluting the useful information.
    \item \textbf{1 --- Wrong topic}: The summary addresses a different regulatory topic than the criterion.
\end{itemize}

\medskip
\textbf{Output:} \texttt{\{"score": <int 1-5>, "reason": "<one sentence>"\}}
\end{tcolorbox}

\begin{tcolorbox}[colback=gray!5, colframe=gray!75, title=Completeness Judge, fonttitle=\bfseries\small, breakable]
\small
You are an expert evaluator for regulatory RAG pipeline outputs.
Assess whether the SUMMARY captures all compliance-critical information from the SOURCE PASSAGES.

\medskip
\textbf{Inputs:} \texttt{\{criterion\_text\}}, \texttt{\{chunk\_texts\}}, \texttt{\{summary\_text\}}, \texttt{\{key\_facts\}}, \texttt{\{thresholds\}}.

\medskip
\textbf{Evaluation Rules:}
\begin{enumerate}[nosep,leftmargin=1.5em]
    \item Identify all compliance-critical elements in the passages: quantitative thresholds, mandatory requirements, conditions, deadlines, and exceptions.
    \item Classify each element as \emph{major} (quantitative thresholds, mandatory requirements, binding conditions) or \emph{minor} (definitions, contextual details, non-binding guidance).
    \item Check whether each element is captured in the summary or key facts/thresholds lists.
    \item A major element omitted from all three outputs (summary, key facts, thresholds) is a substantive gap.
\end{enumerate}

\medskip
\textbf{Scoring Rubric (1--5):}
\begin{itemize}[nosep,leftmargin=*]
    \item \textbf{5 --- Fully complete}: All major and minor compliance-critical elements from the passages are captured in the summary, key facts, or thresholds.
    \item \textbf{4 --- One minor omission}: All major elements captured. Exactly one minor element (a definition, a non-binding recommendation) is omitted.
    \item \textbf{3 --- One major omission}: One major compliance-critical element is missing, or multiple minor elements are omitted.
    \item \textbf{2 --- Multiple major omissions}: Two or more major elements are missing. A compliance assessor would lack critical information.
    \item \textbf{1 --- Mostly incomplete}: The majority of compliance-critical content from the passages is missing from the summary.
\end{itemize}

\medskip
\textbf{Output:} \texttt{\{"score": <int 1-5>, "reason": "<one sentence>"\}}
\end{tcolorbox}

\begin{tcolorbox}[colback=gray!5, colframe=gray!75, title=Coverage Judge, fonttitle=\bfseries\small, breakable]
\small
You are an expert evaluator for regulatory RAG pipeline outputs.
Assess whether the SUMMARY addresses the information needs expressed by the CRITERION.

\medskip
\textbf{Inputs:} \texttt{\{criterion\_text\}}, \texttt{\{chunk\_texts\}}, \texttt{\{summary\_text\}}, \texttt{\{key\_facts\}}, \texttt{\{thresholds\}}.

\medskip
\textbf{Evaluation Rules:}
\begin{enumerate}[nosep,leftmargin=1.5em]
    \item Analyze the criterion to identify its information needs: what specific regulatory content does the criterion reference or require?
    \item Classify each need as \emph{primary} (the specific requirement, threshold, or condition the criterion explicitly references) or \emph{secondary} (supporting context that aids interpretation).
    \item Coverage measures retrieval adequacy: did the pipeline retrieve and summarize passages that address what the criterion asks for?
    \item A criterion referencing a specific article expects the summary to cover that article's content. A criterion referencing a general regulation expects broader coverage of relevant provisions.
\end{enumerate}

\medskip
\textbf{Scoring Rubric (1--5):}
\begin{itemize}[nosep,leftmargin=*]
    \item \textbf{5 --- Full coverage}: All primary and secondary information needs of the criterion are addressed by the summary.
    \item \textbf{4 --- One secondary need unmet}: All primary needs addressed. One secondary information need (supporting context, related provision) is not covered.
    \item \textbf{3 --- Partial primary coverage}: One primary need is only partially addressed, or multiple secondary needs are unmet.
    \item \textbf{2 --- Inadequate coverage}: The primary information need is only partially addressed. A compliance assessor would need to consult additional sources.
    \item \textbf{1 --- No coverage}: The summary does not address the criterion's core information need.
\end{itemize}

\medskip
\textbf{Output:} \texttt{\{"score": <int 1-5>, "reason": "<one sentence>"\}}
\end{tcolorbox}

\section{Supplementary Methodology Details}
\label{app:method_details}

\begin{table}[t]
\centering
\small
\caption{Stages of the \framework{} pipeline.  Each stage receives the cumulative extraction state and the original HTML, applying its specialized competency before passing control to the next stage.}
\label{tab:agents}
\resizebox{.9\columnwidth}{!}{%
\begin{tabular}{@{}clp{4.2cm}@{}}
\toprule
\textbf{Stage} & \textbf{Role} & \textbf{Primary Output} \\
\midrule
$s_1$ & Structural Parser & Criterion hierarchy from HTML \\
$s_2$ & Threshold Extractor & Quantitative \& temporal thresholds \\
$s_3$ & Content Classifier & Category, applicability, eval.\ logic \\
$s_4$ & Reference Extractor & External sources \& CELEX IDs \\
$s_5$ & Dependency Resolver & Inter-criteria relationships \\
$s_6$ & Footnote Processor & Footnote text \& anchor linkage \\
$s_7$ & Schema Assembler & Validated JSON output \\
\bottomrule
\end{tabular}}
\end{table}

\subsection{Semantic Anchoring Procedure}
\label{app:anchoring}

For unnumbered paragraphs that are not resolved by the structural fast-path (\S\ref{subsec:agents}), the LLM performs a three-stage analysis:
\begin{enumerate}[nosep,leftmargin=1.5em]
    \item \textbf{Reference extraction}: the LLM identifies all criterion references within the paragraph (e.g., ``point 1(f)'', ``criteria referred to in 1(b)'') with confidence scores;
    \item \textbf{Scope analysis}: determines whether the paragraph applies to a single criterion, a range, or an entire section, using the extracted references and surrounding context;
    \item \textbf{Hierarchy determination}: assigns the paragraph's position in the hierarchy (sibling vs.\ child of the referenced criterion) based on its semantic function.
\end{enumerate}
A separate \emph{semantic classification} step then applies a three-part reasoning chain (\emph{audit testability}, \emph{obligation direction}, and \emph{compliance impact}) to assign each paragraph a descriptive type such as Verification, Assessment, or BackgroundInformation.
This type is used to generate stable, semantically meaningful identifiers (e.g., \texttt{1(f).Verification}) with deduplication.

\paragraph{Data Acquisition.}
The technical screening criteria for all 242 activities were obtained from the EU Taxonomy
Compass,\footnote{\url{https://ec.europa.eu/sustainable-finance-taxonomy/taxonomy-compass}} the European Commission's official platform for navigating the Taxonomy Regulation.
The platform provides each activity's criteria as raw HTML; all extraction challenges described in
\S\ref{sec:introduction}---implicit hierarchy, unnumbered paragraphs, ambiguous cross-references---are fully present in this input.

  \paragraph{Baseline Selection.}
  We compare against a GPT-4o single-pass baseline rather than
  prior systems (Table~\ref{tab:comparison}) for three reasons:
  (1)~each system targets different regulatory texts with
  incompatible output schemas and extraction granularity,
  making score comparison meaningless;
  (2)~adapting a system designed for a different regulatory
  domain to the EU Taxonomy would require reimplementing its
  extraction logic, risking an uncharitable strawman comparison;
  and (3)~our baseline uses GPT-4o, a frontier commercial LLM,
  which receives the identical input HTML, extraction specification,
  output schema, and few-shot examples in a single call,
  providing a controlled test of whether multi-agent decomposition
  with a smaller model outperforms monolithic generation with a
  stronger one.

\paragraph{Ablation Study.}
To isolate the contribution of each mechanism, we evaluate two
ablated variants: \textbf{$-$ODR}, which removes all
self-correction loops so each stage runs once; and
\textbf{$-$Graph}, which disables cross-stage structural
validation (Table~\ref{tab:ablation}).

\begin{table}[t]
\centering
\small
\caption{Ablation results on $n{=}100$ gold-annotated activities.
Structural and classification metrics are percentages; semantic
equivalence uses a 1--5 GPT-4o judge scale.}
\label{tab:ablation}
\resizebox{\columnwidth}{!}{%
\begin{tabular}{lcccc}
\toprule
\textbf{Metric} & \textbf{GPT-4o} & \textbf{$-$ODR} & \textbf{$-$Graph} & \textbf{Full} \\
\midrule
\multicolumn{5}{l}{\textit{Structural \& Classification (\%)}} \\
Structural F1       & 78.6 & 89.5 & 90.8 & \textbf{94.12} \\
Category Acc.       & 90.2 & 96.3 & 98.1 & \textbf{98.6}  \\
Applicability Acc.  & 85.7 & 93.6 & 96.4 & \textbf{97.24} \\
Eval Logic Acc.     & 80.3 & 89.2 & 87.6 & \textbf{93.4}  \\
\midrule
\multicolumn{5}{l}{\textit{Semantic Equivalence (1--5)}} \\
Threshold           & 3.23 & 4.12 & 4.35 & \textbf{4.43} \\
Reference           & 3.34 & 4.38 & 4.71 & \textbf{4.77} \\
Footnote            & 3.12 & 4.09 & 4.39 & \textbf{4.48} \\
Dependency          & 2.96 & 4.21 & 4.35 & \textbf{4.63} \\
\bottomrule
\end{tabular}}
\end{table}

Removing ODR retains reasonable structural quality (89.5\% F1)
thanks to graph enforcement, but degrades semantic metrics, with Dependency
showing the largest drop (4.63$\to$4.21) as cross-reference
corrections (\S\ref{subsec:agents}, Stage~5) require iterative
analysis across multiple passes.
Removing the graph preserves semantic content (Reference 4.71,
Threshold 4.35) but causes the largest structural drop on
Evaluation Logic (93.4$\to$87.6\%), since the graph enforces
logic--child-count consistency (\S\ref{subsec:graph}) that no
single stage can verify.
Both components contribute complementary gains: ODR refines
within-stage semantic accuracy, while the graph ensures
cross-stage structural coherence.

\subsection{CELEX Identifier Parsing}
\label{app:celex}

EU legislation references are normalized to CELEX identifiers through a four-step parsing chain:
(1)~extract year and document number via format-specific patterns;
(2)~determine document type from contextual keywords (Regulation, Directive, Decision);
(3)~construct the machine-readable identifier (e.g., \texttt{32018R1999});
(4)~validate format and recover from common malformations (transposed digits, missing prefixes).

%

\subsection{Structural F1 Computation}
\label{app:structural_f1}

Structural F1 measures how well a system-produced criterion tree matches the gold-standard tree. The computation proceeds in two stages.

\paragraph{Node Alignment.}
Both trees are flattened into node lists, and nodes are paired through a two-pass process. The first pass matches nodes by exact \texttt{criterion\_id}. The second pass takes all remaining unmatched nodes and pairs them greedily by verbatim text similarity, accepting any pair whose \texttt{SequenceMatcher} ratio reaches at least 0.80. This produces three sets: matched pairs, unmatched gold nodes (false negatives), and unmatched system nodes (false positives).

\paragraph{Context Quality Score.}
Each matched pair receives a quality score between 0 and 1, computed as a weighted sum of four features:
\begin{enumerate}[nosep,leftmargin=1.5em]
    \item \textbf{Parent placement} (weight 0.30): whether the system node sits under the same parent \texttt{criterion\_id} as the gold node. A correctly identified node placed under the wrong parent receives no credit for this feature.
    \item \textbf{Sibling overlap} (weight 0.25): Jaccard similarity between the sibling ID sets of the system and gold nodes. If gold groups criteria \{a, b, c\} under one parent but the system groups \{a, b, d\}, partial credit reflects the overlap.
    \item \textbf{Subtree shape} (weight 0.25): ratio of child counts, computed as $\min(s, g) / \max(s, g)$ where $s$ and $g$ are the number of children in the system and gold trees. A leaf node matched to another leaf receives full credit.
    \item \textbf{Schema completeness} (weight 0.20): fraction of the 13 required output fields present as keys on the system node.  A node missing two of thirteen fields receives $11/13 \approx 0.85$ for this feature.
\end{enumerate}

\paragraph{F1 Computation.}
The quality scores are summed across all matched pairs to produce a weighted true-positive count. Precision divides this count by the total number of system nodes (matched plus unmatched), and recall divides it by the total number of gold nodes (matched plus unmatched). F1 is the harmonic mean of precision and recall.

\subsection{RAG Query Refinement}
\label{app:rag_eval}

Figure~\ref{fig:query_refinement} illustrates the ReAct loop's query refinement across iterations.
Queries become progressively more specific, targeting particular articles and provisions.

\begin{figure*}[t]
\begin{tcolorbox}[colback=white,colframe=black!60,boxrule=0.4pt,arc=2pt,left=4pt,right=4pt,top=4pt,bottom=4pt,title={\small Query Refinement Examples (3-Iteration Cases)}]
\small

\textbf{Example 1:} Biodiversity --- Habitats Directive (92/43/EEC)\\
Final retrieval confidence: \textbf{0.90}

\medskip
\textit{Iteration 1:} ``What are the conditions for classification of protected areas under the IUCN system and Natura 2000 sites according to Directive 92/43/EEC?''

\smallskip
\textit{Iteration 3:} ``What specific legal provisions does Directive 92/43/EEC establish for classifying protected areas under the EU Taxonomy?''

\bigskip
\textbf{Example 2:} Afforestation --- EU Taxonomy (2020/852)\\
Final retrieval confidence: \textbf{0.48}

\medskip
\textit{Iteration 1:} ``What does Article 11(1) of the EU Taxonomy define as substantial contribution for afforestation under climate change adaptation?''

\smallskip
\textit{Iteration 3:} ``What are the specific criteria and thresholds under Article 11(1) for demonstrating substantial contribution in afforestation?''
\end{tcolorbox}
\caption{Query refinement across ReAct iterations. Example~1 achieves high confidence after narrowing scope to specific provisions. Example~2 correctly assigns low confidence when the target article lacks retrievable thresholds.}
\label{fig:query_refinement}
\end{figure*}

\section{Output Schema}
\label{app:schema}

Each criterion node in the \dataset{} dataset conforms to the following fixed 13-field JSON schema.
Complex objects (\texttt{threshold}, \texttt{references}, \texttt{dependencies}) are set to \texttt{null} when they contain no meaningful data; \texttt{footnotes} is always an array (\texttt{[]} when empty, never \texttt{null}).

\onecolumn
\begin{tcolorbox}[
    title={\textbf{Fixed 13-Field Criterion Schema}},
    colback=white,colframe=black!70,fonttitle=\small,
    boxrule=0.4pt,arc=2pt,left=3pt,right=3pt,top=2pt,bottom=2pt,
    width=\textwidth,
    breakable
]
\begin{lstlisting}[language=json, basicstyle=\ttfamily\fontsize{7pt}{8.6pt}\selectfont, xleftmargin=0pt]
{
  "criterion_id": "",
  "category": "Quantitative" | "Qualitative",
  "applicability": "Mandatory" | "Conditional",
  "tags": ["Verification", "Methodology", "Commitment",
           "Assessment", "Replacement",
           "BackgroundInformation"],
  "verbatim_text": "",
  "rule_summary": "",
  "evaluation_logic": "AND" | "OR" | "N_OF_K" | "LEAF",
  "n_required": null,
  "threshold": null | {
    "quantitative": null | {
      "logic": "AND" | "OR" | null,
      "items": [
        {
          "metric": "",
          "operator": ">" | "<" | ">=" | "<=" | "=",
          "value": 0,
          "unit": "",
          "period": null | {
            "type": "bounded",
            "value": 0,
            "unit": "years" | "months" | "days"
          } | {
            "type": "lifetime"
          }
        }
      ]
    },
    "temporal": null | {
      "logic": "AND" | "OR" | null,
      "items": [
        { "type": "deadline", "date": "YYYY-MM-DD" } |
        { "type": "effective_from",
          "date": "YYYY-MM-DD" } |
        { "type": "effective_until",
          "date": "YYYY-MM-DD" } |
        { "type": "window",
          "start": "YYYY-MM-DD",
          "end": "YYYY-MM-DD" } |
        { "type": "recurring",
          "interval_value": 0,
          "interval_unit": "years" | "months" | "days" }
      ]
    }
  },
  "references": null | {
    "logic": "AND" | "OR" | null,
    "sources": [
      {
        "text": "",
        "type": "citation_only" | "must_be_fetched",
        "link": "",
        "link_status": "provided" | "generated",
        "celex_id": null | "32018L2001",
        "enrichment": null | {
          "status": "retrieved"|"paywalled"|"skipped",
          "summary": "",
          "key_facts": [],
          "thresholds": [],
          "confidence": 0.0
        }
      }
    ]
  },
  "dependencies": null | {
    "condition_summary": "",
    "min_conditions_to_meet": 1,
    "clauses": [
      { "criterion_id": "", "status": "Affirmation" | "Negation" }
    ]
  },
  "sub_criteria": [],
  "footnotes": [
    {
      "id": "fn-1",
      "categories": [],
      "items": [
        {
          "kind": "EU Legal Act" | "Standard" | "...",
          "title": "",
          "celex_id": null | "32018L2001",
          "type": "citation_only" | "must_be_fetched",
          "oj": null | "",
          "enrichment": null | {
            "status": "retrieved"|"paywalled"|"skipped",
            "summary": "",
            "key_facts": [],
            "thresholds": [],
            "confidence": 0.0
          }
        }
      ],
      "definitions": [],
      "notes": []
    }
  ]
}
\end{lstlisting}
\end{tcolorbox}
\captionof{figure}{Fixed 13-field JSON schema for each criterion node in \dataset{}. The \texttt{enrichment} sub-object inside each reference source and footnote item is populated by the RAG sub-pipeline (\S\ref{subsec:rag}); fields \texttt{summary} through \texttt{confidence} are present only when \texttt{status\,=\,"retrieved"}. See Figure~\ref{fig:json_output_1} for a populated example.}
\label{fig:schema}
\twocolumn

\section{Gold Annotation Details}
\label{app:annotation}




The 100 gold-annotated activities were selected via stratified sampling across three structural complexity tiers:
\begin{itemize}[nosep,leftmargin=1.5em]
    \item \textbf{Simple} (30 activities): leaf-node criteria with a single threshold and no inter-criteria dependencies.
    \item \textbf{Medium} (40 activities): a chapeau node with 2--5 children connected by basic \textsc{and}/\textsc{or} evaluation logic.
    \item \textbf{Complex} (30 activities): transitional pathways, ancillary nodes, footnotes, cross-reference corrections, and threshold inheritance.
\end{itemize}
Activities were sampled so that all three Delegated Acts are represented, capturing different drafting conventions.

The extraction pipeline generated candidate outputs for each activity.
Two PhD students, both trained on the extraction specification (Appendix~\ref{app:schema}), served as annotators.
Each annotator independently reviewed 50 of the 100 activities across nine dimensions: applicability, category, evaluation logic, tags, thresholds, references, dependencies, footnotes, and structure.
The two disjoint sets were assigned so that each complexity tier and each Delegated Act were represented in both.

To measure annotation reliability, a subset of 15 activities (5 per complexity tier) was drawn from the 100 and re-annotated by the annotator who had not been assigned it, giving these 15 two independent annotations.
Table~\ref{tab:iaa} reports Cohen's $\kappa$ for the three core evaluation dimensions.

\begin{table}[h]
\centering
\small
\caption{Inter-annotator agreement on the shared 15-activity subset.}
\label{tab:iaa}
\resizebox{.8\columnwidth}{!}{%
\begin{tabular}{@{}lc@{}}
\toprule
\textbf{Dimension} & \textbf{Cohen's $\kappa$} \\
\midrule
Hierarchical structure & 0.91 \\
Dependencies (incl.\ cross-reference corrections) & 0.86 \\
Threshold extraction & 0.84 \\
\bottomrule
\end{tabular}}
\end{table}

Disagreements on the shared subset were resolved through discussion until consensus was reached.
The resulting corrections were applied to produce the final gold annotations used throughout the evaluation.

\paragraph{Judge Validation.}
To validate LLM-judge reliability, one annotator independently
re-scored a stratified random subset of 50 criterion--dimension pairs
using the same rubrics and 1--5 scales: 30 pairs drawn from both
\framework{} and baseline outputs across the four semantic
equivalence dimensions, and 20 pairs from \framework{} outputs
across the four RAG quality dimensions. Human--judge agreement yielded Spearman $\rho{=}0.86$ overall ($0.87$ for semantic equivalence, $0.85$ for RAG quality),
with 90\% of scores (45/50) falling within $\pm 1$ point and Cohen's $\kappa{=}0.83$ on binarized ratings ($\leq 3$ vs.\ $\geq 4$).

\end{document}